\documentclass[11pt]{article}
\usepackage{amsmath,amsfonts,cite}

\newcommand{\blockA}[3]{\ensuremath{
     \setlength{\unitlength}{1mm}
     \begin{picture}(10,15)(0,0)
          \put( 0, 4){\line(1,0){10}}
          \put(10, 4){\line(0,1){10}}
          \put( 2, 5){\makebox(0,0)[lb]{\small $#1$}}
          \put(11,13){\makebox(0,0)[lt]{\small $#2$}}
          \put(10, 3){\makebox(0,0)[ct]{\small $#3$}}
     \end{picture}}}

\newcommand{\blockB}[1]{\ensuremath{
     \setlength{\unitlength}{1mm}
     \begin{picture}(10,15)(0,0)
          \put( 0, 4){\line(1,0){10}}
          \put( 8, 5){\makebox(0,0)[rb]{\small $#1$}}
     \end{picture}}}

\newcommand{\blockC}[3]{\ensuremath{
     \setlength{\unitlength}{1mm}
     \begin{picture}(10,15)(0,0)
          \put( 0, 4){\line(1,0){10}}
          \put(10, 4){\line(0,1){10}}
          \put( 5, 5){\makebox(0,0)[cb]{\small $#1$}}
          \put(11,13){\makebox(0,0)[lt]{\small $#2$}}
          \put(10, 3){\makebox(0,0)[ct]{\small $#3$}}
     \end{picture}}}

\newcommand{\blockD}[3]{\ensuremath{
     \setlength{\unitlength}{1mm}
     \begin{picture}(8,15)(0,0)
          \put( 0, 4){\line(1,0){8}}
          \put( 8, 4){\line(0,1){10}}
          \put( 1, 5){\makebox(0,0)[cb]{\small $#1$}}
          \put( 9,13){\makebox(0,0)[lt]{\small $#2$}}
          \put( 8, 3){\makebox(0,0)[ct]{\small $#3$}}
     \end{picture}}}

\newcommand{\blockE}[0]{\ensuremath{
     \setlength{\unitlength}{1mm}
     \begin{picture}(3,15)(0,0)
          \put( 0, 4){\line(1,0){3}}
     \end{picture}}}

\newcommand{\blockUp}[1]{\ensuremath{
     \setlength{\unitlength}{1mm}
     \put( 0,10){#1} }}

\newcommand{\cbA}[4]{\blockA{#1}{#2}{#4}\blockB{#3}}
\newcommand{\cbB}[7]{\blockA{#1}{#2}{#6}\blockC{#5}{#3}{#7}\blockB{#4}}
\newcommand{\cbC}[7]{\blockA{#1}{}{#6}%
     \blockUp{\blockD{#5}{#2}{#7}\blockB{#3}}%
     \blockE\blockB{#4}}

\newcommand{\bL}[1]{\raisebox{-3mm}{#1}}

\newcommand{\F}[6]{\mbox{\ensuremath{
     \textsf{\large F}_{\raisebox{-1pt}{\scriptsize\!\!$#5#6$}}\!\left[\begin{array}{ll}
     \!#2 & \!#3\! \\
     \!#1 & \!#4\!
     \end{array}\!\right]}}}

\newcommand{\Bmat}[7]{\mbox{\ensuremath{
     \textsf{\large B}^{(#1)}_{\raisebox{-1pt}{\scriptsize\!\!$#6#7$}}\!\left[\begin{array}{ll}
     \!#3 & \!#4\! \\
     \!#2 & \!#5\!
     \end{array}\!\right]}}}

\newcommand{\B}[3]{\ensuremath{\,{^{(\!#1\!)}}{\!B_{#2}}^{#3}\,}}
\newcommand{\bc}[6]{\ensuremath{\,C^{(\!#1\hspace{-0.3pt}#2\hspace{-0.3pt}#3\!)#6}_{\;#4#5}\,}}
\newcommand{\bnpt}[3]{\ensuremath{\langle{#1}\rangle^{\!(\!#2\!)}_{\text{\tiny{#3}}}}}
\newcommand{\bp}[3]{\ensuremath{\,\psi^{(\!#1\hspace{-0.7pt}#2\!)}_{#3}}}
\newcommand{\C}[3]{\ensuremath{\,{C_{{#1}{#2}}}^{#3}\,}}

\newcommand{\eps}{\ensuremath{\epsilon}}

\newcommand{\hb}{\ensuremath{\bar h}}
\newcommand{\ib}{\ensuremath{\bar\imath}}
\newcommand{\is}[1]{\ensuremath{|#1\rangle}}

\newcommand{\jb}{\ensuremath{\bar\jmath}}
\renewcommand{\l}{\ensuremath{\ell}}

\newcommand{\N}[3]{\ensuremath{{\mathrm{N}_{#1#2}}^{#3}}}
\newcommand{\n}[3]{\ensuremath{{n_{#1#2}}^{#3}}}
\newcommand{\npt}[1]{\ensuremath{\langle{#1}\rangle}}

\renewcommand{\o}{\ensuremath{\omega}}
\newcommand{\os}[1]{\ensuremath{\langle{#1}|}}
\newcommand{\ph}[1]{\ensuremath{\phi_{#1}}}

\renewcommand{\S}[2]{\ensuremath{{S_{#1}}^{#2}}}

\newcommand{\simuu}[4]{\ensuremath{\underset{#3\rightarrow #4}{\underset{#1\rightarrow #2}{\sim}}}}

\newcommand{\wb}{\ensuremath{\bar w}}
\newcommand{\Z}{\ensuremath{\mathbb{Z}}}
\newcommand{\zb}{\ensuremath{\bar z}}

\newcommand{\bfb}[3]{
     \raisebox{-1.45mm}{
     \setlength{\unitlength}{0.85mm}
     \begin{picture}(20,5)(2,0)
          \put(   0, 1){\line(1,0){9.3}}
          \put(10.7, 1){\line(1,0){9.3}}
          \put(  10, 1){\qbezier(-0.7,0)(-0.7,0.7)(0,0.7)\qbezier(0,0.7)(0.7,0.7)(0.7,0)}
          \put(   4, 1.8){\makebox(0,0)[cb]{\small\ensuremath{\smash[b]{#1}}}}
          \put(  16, 1.8){\makebox(0,0)[cb]{\small\ensuremath{\smash[b]{#3}}}}
          \put(  10, 2.7){\makebox(0,0)[cb]{\small\ensuremath{\smash[b]{#2}}}}
     \end{picture}}}

\newcommand{\bfbfb}[5]{
     \raisebox{-1.45mm}{
     \setlength{\unitlength}{0.85mm}
     \begin{picture}(30,5)(2,0)
          \put(   0, 1){\line(1,0){9.3}}
          \put(10.7, 1){\line(1,0){8.6}}
          \put(20.7, 1){\line(1,0){9.3}}
          \put(10, 1){\qbezier(-0.7,0)(-0.7,0.7)(0,0.7)\qbezier(0,0.7)(0.7,0.7)(0.7,0)}
          \put(20, 1){\qbezier(-0.7,0)(-0.7,0.7)(0,0.7)\qbezier(0,0.7)(0.7,0.7)(0.7,0)}
          \put( 4, 1.8){\makebox(0,0)[cb]{\small \ensuremath{\smash[b]{#1}}}}
          \put(15, 1.8){\makebox(0,0)[cb]{\small \ensuremath{\smash[b]{#3}}}}
          \put(26, 1.8){\makebox(0,0)[cb]{\small \ensuremath{\smash[b]{#5}}}}
          \put(10, 2.7){\makebox(0,0)[cb]{\small \ensuremath{\smash[b]{#2}}}}
          \put(20, 2.7){\makebox(0,0)[cb]{\small \ensuremath{\smash[b]{#4}}}}
     \end{picture}}}

\newcommand{\bfbfbfb}[7]{
     \raisebox{-1.45mm}{
     \setlength{\unitlength}{0.85mm}
     \begin{picture}(40,5)(2,0)
          \put(   0, 1){\line(1,0){9.3}}
          \put(10.7, 1){\line(1,0){8.6}}
          \put(20.7, 1){\line(1,0){8.6}}
          \put(30.7, 1){\line(1,0){9.3}}
          \put(10, 1){\qbezier(-0.7,0)(-0.7,0.7)(0,0.7)\qbezier(0,0.7)(0.7,0.7)(0.7,0)}
          \put(20, 1){\qbezier(-0.7,0)(-0.7,0.7)(0,0.7)\qbezier(0,0.7)(0.7,0.7)(0.7,0)}
          \put(30, 1){\qbezier(-0.7,0)(-0.7,0.7)(0,0.7)\qbezier(0,0.7)(0.7,0.7)(0.7,0)}
          \put( 4, 1.8){\makebox(0,0)[cb]{\small \ensuremath{\smash[b]{#1}}}}
          \put(15, 1.8){\makebox(0,0)[cb]{\small \ensuremath{\smash[b]{#3}}}}
          \put(25, 1.8){\makebox(0,0)[cb]{\small \ensuremath{\smash[b]{#5}}}}
          \put(36, 1.8){\makebox(0,0)[cb]{\small \ensuremath{\smash[b]{#7}}}}
          \put(10, 2.7){\makebox(0,0)[cb]{\small \ensuremath{\smash[b]{#2}}}}
          \put(20, 2.7){\makebox(0,0)[cb]{\small \ensuremath{\smash[b]{#4}}}}
          \put(30, 2.7){\makebox(0,0)[cb]{\small \ensuremath{\smash[b]{#6}}}}
     \end{picture}}}

\newcommand{\bfbfbfbfb}[9]{
     \raisebox{-1.45mm}{
     \setlength{\unitlength}{0.85mm}
     \begin{picture}(50,5)(2,0)
          \put(   0, 1){\line(1,0){9.3}}
          \put(10.7, 1){\line(1,0){8.6}}
          \put(20.7, 1){\line(1,0){8.6}}
          \put(30.7, 1){\line(1,0){8.6}}
          \put(40.7, 1){\line(1,0){9.3}}
          \put(10, 1){\qbezier(-0.7,0)(-0.7,0.7)(0,0.7)\qbezier(0,0.7)(0.7,0.7)(0.7,0)}
          \put(20, 1){\qbezier(-0.7,0)(-0.7,0.7)(0,0.7)\qbezier(0,0.7)(0.7,0.7)(0.7,0)}
          \put(30, 1){\qbezier(-0.7,0)(-0.7,0.7)(0,0.7)\qbezier(0,0.7)(0.7,0.7)(0.7,0)}
          \put(40, 1){\qbezier(-0.7,0)(-0.7,0.7)(0,0.7)\qbezier(0,0.7)(0.7,0.7)(0.7,0)}
          \put( 4, 1.8){\makebox(0,0)[cb]{\small \ensuremath{\smash[b]{#1}}}}
          \put(15, 1.8){\makebox(0,0)[cb]{\small \ensuremath{\smash[b]{#3}}}}
          \put(25, 1.8){\makebox(0,0)[cb]{\small \ensuremath{\smash[b]{#5}}}}
          \put(35, 1.8){\makebox(0,0)[cb]{\small \ensuremath{\smash[b]{#7}}}}
          \put(46, 1.8){\makebox(0,0)[cb]{\small \ensuremath{\smash[b]{#9}}}}
          \put(10, 2.7){\makebox(0,0)[cb]{\small \ensuremath{\smash[b]{#2}}}}
          \put(20, 2.7){\makebox(0,0)[cb]{\small \ensuremath{\smash[b]{#4}}}}
          \put(30, 2.7){\makebox(0,0)[cb]{\small \ensuremath{\smash[b]{#6}}}}
          \put(40, 2.7){\makebox(0,0)[cb]{\small \ensuremath{\smash[b]{#8}}}}
     \end{picture}}}

\newcommand{\AdiagB}[1]{
     \raisebox{-2mm}{
     \setlength{\unitlength}{0.7mm}
     \begin{picture}(50,10)(0,0)
          \put( 2, 3){\circle{5}}
          \put( 2, 3){\circle*{2}}
          \put( 5, 5){\makebox(0,0)[cb]{\small 1}}
          \put( 2, 3){\line(1,0){8}}
          \put(10, 3){\circle*{2}}
          \put(11, 5){\makebox(0,0)[cb]{\small 2}}
          \put(10, 3){\line(1,0){8}}
          \put(18, 3){\circle{5}}
          \put(18, 3){\circle*{2}}
          \put(18, 3){\line(1,0){3}}
          \put(22, 3){\makebox(0,0)[lc]{\small $\cdots$}}
          \put(29, 3){\line(1,0){3}}
          \put(33, 3){\circle{5}}
          \put(33, 3){\circle*{2}}
          \put(33, 3){\line(1,0){8}}
          \put(41, 3){\circle*{2}}
          \put(41, 5){\makebox(0,0)[lb]{\small $#1{-}1$}}
     \end{picture}}}

\newcommand{\AdiagC}[1]{
     \raisebox{-2mm}{
     \setlength{\unitlength}{0.7mm}
     \begin{picture}(50,10)(0,0)
          \put( 2, 3){\circle{5}}
          \put( 2, 3){\circle*{2}}
          \put( 5, 5){\makebox(0,0)[cb]{\small 1}}
          \put( 2, 3){\line(1,0){8}}
          \put(10, 3){\circle{5}}
          \put(10, 3){\circle*{2}}
          \put(13, 5){\makebox(0,0)[cb]{\small 2}}
          \put(10, 3){\line(1,0){8}}
          \put(18, 3){\circle{5}}
          \put(18, 3){\circle*{2}}
          \put(18, 3){\line(1,0){3}}
          \put(22, 3){\makebox(0,0)[lc]{\small $\cdots$}}
          \put(29, 3){\line(1,0){3}}
          \put(33, 3){\circle{5}}
          \put(33, 3){\circle*{2}}
          \put(33, 3){\line(1,0){8}}
          \put(41, 3){\circle{5}}
          \put(41, 3){\circle*{2}}
          \put(42, 5){\makebox(0,0)[lb]{\small $#1{-}1$}}
     \end{picture}}}

\newcommand{\DdiagB}[1]{
     \raisebox{-6.6mm}{
     \setlength{\unitlength}{0.7mm}
     \begin{picture}(55,20)(0,0)
          \put( 2,10){\circle{5}}
          \put( 2,10){\circle*{2}}
          \put( 5,12){\makebox(0,0)[cb]{\small 1}}
          \put( 2,10){\line(1,0){8}}
          \put(10,10){\circle{5}}
          \put(10,10){\circle*{2}}
          \put(13,12){\makebox(0,0)[cb]{\small 2}}
          \put(10,10){\line(1,0){8}}
          \put(18,10){\circle{5}}
          \put(18,10){\circle*{2}}
          \put(18,10){\line(1,0){3}}
          \put(22,10){\makebox(0,0)[lc]{\small $\cdots$}}
          \put(29,10){\line(1,0){3}}
          \put(33,10){\line(1,1){6}}
          \put(33,10){\line(1,-1){6}}
          \put(33,10){\circle{5}}
          \put(33,10){\circle*{2}}
          \put(39,16){\circle{5}}
          \put(39,16){\circle*{2}}
          \put(39, 4){\circle{5}}
          \put(39, 4){\circle*{2}}
          \put(44,16){\makebox(0,0)[lc]{\small $\frac{#1}{2}$}}
          \put(44, 4){\makebox(0,0)[lc]{\small $\frac{#1}{2}{+}1$}}
     \end{picture}}}

\newcommand{\DdiagC}{
     \raisebox{-6.6mm}{
     \setlength{\unitlength}{0.7mm}
     \begin{picture}(45,20)(0,0)
          \put( 2,10){\circle{5}}
          \put( 2,10){\circle*{2}}
          \put( 2,10){\line(1,0){8}}
          \put(10,10){\circle{5}}
          \put(10,10){\circle*{2}}
          \put(10,10){\line(1,0){8}}
          \put(18,10){\circle{5}}
          \put(18,10){\circle*{2}}
          \put(18,10){\line(1,0){3}}
          \put(22,10){\makebox(0,0)[lc]{\small $\cdots$}}
          \put(29,10){\line(1,0){3}}
          \put(33,10){\line(1,1){6}}
          \put(33,10){\line(1,-1){6}}
          \put(33,10){\circle{5}}
          \put(33,10){\circle*{2}}
          \put(39,16){\circle*{2}}
          \put(39, 4){\circle*{2}}
     \end{picture}}}

\newcommand{\DdiagD}{
     \raisebox{-6.6mm}{
     \setlength{\unitlength}{0.7mm}
     \begin{picture}(45,20)(0,0)
          \put( 2,10){\circle*{2}}
          \put( 2,10){\line(1,0){8}}
          \put(10,10){\circle*{2}}
          \put(10,10){\line(1,0){8}}
          \put(18,10){\circle*{2}}
          \put(18,10){\line(1,0){3}}
          \put(22,10){\makebox(0,0)[lc]{\small $\cdots$}}
          \put(29,10){\line(1,0){3}}
          \put(33,10){\line(1,1){6}}
          \put(33,10){\line(1,-1){6}}
          \put(33,10){\circle*{2}}
          \put(39,16){\circle{5}}
          \put(39,16){\circle*{2}}
          \put(39, 4){\circle{5}}
          \put(39, 4){\circle*{2}}
     \end{picture}}}

\begin{document}
\setcounter{footnote}{0}
\begin{titlepage}
\vskip 0.5cm
\begin{flushright}
{\tt hep-th/9908046}\\
August 1999 
\end{flushright}
\vskip 1.2cm
\begin{center}
{\Large {\bf Structure constants}} \\[5pt]
{\Large {\bf for the D-series Virasoro minimal models} }
\end{center}
\vskip 0.8cm
\centerline{Ingo Runkel%
\footnote{e-mail: {\tt ingo@mth.kcl.ac.uk}}
}
\vskip 0.6cm
\centerline{\sl Mathematics Department, }
\centerline{\sl King's College London, Strand, London WC2R 2LS, U.K.}
\vskip 0.9cm
\begin{abstract}
\vskip0.15cm
\noindent 
In this paper expressions are given for the bulk and boundary
structure constants of D-series Virasoro minimal
models on the upper half 
plane. It is the continuation of an earlier work on the A-series. The
solution for the boundary theory is found first and then extended to
the bulk. The modular invariant bulk field content is recovered as the
maximal set of bulk fields consistent with the boundary theory. It is
found that the
structure constants are unique up to redefinition of the fields and in
the chosen normalisation exhibit a manifest $\Z_2$--symmetry
associated to the D-diagram. The solution has been subjected to random
numerical tests against the constraints it has to fulfill.
\end{abstract}
\end{titlepage}
\setcounter{footnote}{0}

\section{Introduction}

One of the aims of studying quantum field theories is to compute
correlation functions of the fields in the theory. In two-dimensional
conformal field theory one uses the presence of an infinite amount of
symmetries to achieve this goal.

We will consider a special subclass of conformal
field theories, called minimal models. One of their properties is that
the correlators fulfill certain linear
differential equations (see \cite{BPZ84} or e.g.~\cite{YBk} for
details). One can find a distinguished set of solutions to these
differential equations, called conformal blocks. A correlator
can then be expressed as a bilinear combination of these blocks. To
know the coefficients in this bilinear combination a set of complex
numbers, the structure constants, are needed. Structure constants 
appear in the short distance expansion (or operator
product expansion, OPE) of certain fields in the conformal field
theory, called primary fields.

In this paper we consider only Virasoro minimal models.
One approach to calculate all structure constants is to start in a
situation without boundaries. In this case only bulk fields are
present and one can solve the bulk theory alone by determining the
conformal blocks and a consistent set of structure constants for the
bulk fields (see \cite{DF84} for the spinless case). It is possible to classify all
modular invariant minimal models without boundaries
\cite{CIZ87} with the result that they fall into an
A--D--E pattern. 

Starting from a well defined bulk theory one can proceed and introduce
a boundary into the system. As it turns out the differential equations
that the correlators fulfill stay the same, but now the correlator is
a linear, not a bilinear, combination of the conformal blocks
mentioned above (see \cite{Car84}) and furthermore two new sets of structure constants
appear: In addition to the bulk structure
constants that determine the OPE of two bulk primary fields, there are
the bulk-boundary couplings that describe the expansion of a bulk field in
terms of boundary fields, and the boundary structure constants for the OPE
of two boundary fields.
 
One can now classify all conformally invariant boundary conditions and
the boundary fields that live on these boundaries or interpolate
between two adjacent boundary conditions (see
\cite{CaL91,PSS95,FuS97,BPZ98}). Again one finds an A--D--E pattern
for all possible boundary theories \cite{BPZ98}, plus tadpole
diagrams, which have been discarded as the resulting theory does not match
with any of the modular invariant bulk theories.

All three sets of structure constants can in principle be worked out
from six consistency conditions that arise from taking different
limits in certain correlators and equating the different OPEs in these
limits. The constraints that arise in this way are called sewing
constraints (see \cite{Lew92} and for an extension to non-orientable
surfaces \cite{PSS95}). A more general discussion of the sewing constraints 
in  presence of nontrivial multiplicities and their relation to
\cite{MSb90} can be found in \cite{BPZ99}.

Above we have started from a bulk theory, which can be solved
consistently, and then introduced a boundary. Thus the bulk
structure constants are known first and then a solution for
the other two sets of structure constants consistent with these has to
be obtained. Whereas the bulk structure constants for diagonal and
nondiagonal minimal models can be found in the literature 
\cite{DF84,StrXX,PeZ95},
these calculations have not been carried out for the
boundary structure constants and only for a subset of the
bulk-boundary couplings (see \cite{CaL91,PSS95,FuS97,BPZ98}).

In this paper and a
previous work \cite{Run99} the missing structure constants for the A--
and D-series are calculated by taking the opposite path. The starting
point now is a theory with only boundary fields. One chooses an
A--D--E type boundary field content and tries to determine
a consistent set of boundary structure constants. Once this is
achieved the boundary theory is extended to the bulk and it is only at
this point that bulk fields are introduced. It is interesting that
one finds that the maximal bulk
field content consistent with the boundary theory coincides with the
modular invariant one given in \cite{CIZ87}. 
The bulk-boundary couplings and the bulk structure constants can now
be obtained from the sewing constraints in a straight forward way.
The results are
explicit expressions for all three sets of structure constants 
in terms of so called fusion (F--) matrices (see Appendix~A 
for details and references). The expressions are free
of sign ambiguities (e.g.\ they do not contain any square roots) so
that they are well suited for numerical computations.

The $D_{\text{even}}$ minimal models have a larger symmetry algebra (a
W--algebra, see~\cite{WalXX} and references therein) that contains the
Virasoro algebra as subalgebra. In principle one could understand these
conformal field theories in terms of the W-algebra, with
W-algebra chiral blocks and F--matrices. But since these CFTs are still
finitely reducible with respect to the Virasoro algebra alone
(i.e.\ only a finite number of irreducible highest weight
representations of the Virasoro algebra occur), it is possible and
indeed simple to use only the Virasoro symmetry.

The fact that one does not use the fully extended chiral algebra and
the possible presence of multiple copies of primary fields means that
the sewing constraints in Ref.~\cite{Lew92} must be rederived, which
has been done. The results are presented in
section~\ref{Genus-zero-sewing-constraints} and an explicit 
example is provided in Appendix~A.

Both, in the A-- and the D--series, the sewing constraints
are overdetermined and only a subset is used to find the structure
constants. It has to some extend been tested numerically (but not
proven analytically) that 
the given structure constants solve the full set of constraints. It is
shown that, given the boundary field content, any solution can be brought
to a standard form by redefining the fields. In other words, under
gauge transformations the space of solutions to the sewing constraints
consists of only one orbit.

The formulae for all bulk and boundary structure constants are given in
eqns.~\eqref{collection-first-bsc-formula}--\eqref{collection-last-bsc-formula} 
and \eqref{bulk-field-normalisation}--\eqref{general-C-formula}. A short
description of how to implement them can be found in the
conclusion. The derivation of these expressions is organised as follows:

In section~\ref{Genus-zero-sewing-constraints} the notation for the
various fields and structure constants is introduced and the rederived
sewing constraints for genus zero surfaces are stated in a
suitable form. Section~\ref{Cylinder-partition-function} gives a short
description of 
the procedure introduced in~\cite{BPZ98} for the construction of
cylinder partition functions. This defines the boundary field content
which is used as input in the construction of the solution to the
sewing constraints. The boundary structure 
constants are found in section~\ref{Boundary-structure-constants}.
Finally the extension to the bulk theory is carried out in
section~\ref{Extension-to-bulk-theory}. Here the maximal
bulk field content consistent with the boundary theory is determined
and it is observed to be modular invariant. The remaining
structure constants, i.e.\ the bulk-boundary couplings and the bulk
structure constants are calculated and expressions for the vacuum
expectation value of the unit disc are obtained. A choice
of basis is presented which renders all structure constants
real. Finally it is observed that a subset of all structure constants
is left invariant under a large number of distinct $\Z_2$--actions.

\section{Genus zero sewing constraints}
\label{Genus-zero-sewing-constraints}

Unless otherwise mentioned, throughout this paper the boundary
conformal field theory will be considered on the upper half plane
(UHP) with the boundary given by the real line. 

In a general Virasoro minimal model there can be several primary bulk
or boundary fields transforming in the same highest weight
representation(s) of the Virasoro algebra. Therefore we
cannot label the fields solely by their representations.
Virasoro highest weight representations are labelled
by small letters $i,j,k$ and to take care of fields with multiplicities
these carry an additional Greek index $i_\alpha$, $j_\beta$,
$k_\gamma$. Boundary conditions are labelled by
small letters $a,b,c$ or also $x,y,z$.

It is always possible to make the (bulk-- and boundary--)
two-point functions diagonal by an appropriate choice of labels and
normalisation of the fields:
$\npt{\varphi_{i_\alpha}(x)\varphi_{j_\beta}(y)}{=}
\delta_{i,j}\delta_{\alpha,\beta}\cdot f(x,y)$.

A bulk field $\ph{i_\alpha}$ transforms in the tensor product of two Virasoro
highest weight representations $i{\otimes}\ib$. The two representations have 
conformal weights $h_i, h_{\ib}$ or, for better
readability, $h_i, \hb_i$. A boundary field $\bp ab{k_\alpha}$ lives
between boundary conditions $a$ and $b$ and 
transforms in one Virasoro representation $k$ of conformal weight $h_k$. 

The vacuum expectation value of the UHP with boundary condition $a$
imposed on the real line will be denoted with $\bnpt 1a{UHP}$. The
vacuum expectation value of the full complex plane with no boundaries
present will be denoted with $\npt 1$ or $\npt{0|0}$.

The leading terms in the bulk--bulk, bulk--boundary and
boundary--boundary operator product 
expansions of primary fields are, in this order:
{\allowdisplaybreaks
\begin{align}
     \ph {i_\alpha}(z)\ph {j_\beta}(w) &=
     \sum_{k,\gamma} \C{i_\alpha}{j_\beta}{k_\gamma}  
     (z-w)^{h_k-h_i-h_j}(\zb-\wb)^{\hb_k-\hb_i-\hb_j} \notag\\[-4mm]
     & \qquad\qquad \cdot \left(\ph {k_\gamma}(w) + \cdots \right)
     &&|z|{>}|w|
     \label{OPE-definitions-first}\\
     \ph {i_\alpha}(x+iy) &=
     \sum_{k,\gamma} \B a{i_\alpha}{k_\gamma} \cdot 
     (2y)^{h_k-h_i-\hb_i}\cdot \left(\bp aa{k_\gamma}(x) + \cdots \right)
     && y>0 \\ 
     \bp ab{i_\alpha}(x) \bp bc{j_\beta}(y) &= 
     \sum_{k,\gamma} \bc abc{i_\alpha}{j_\beta}{k_\gamma}\cdot  
     (x-y)^{h_k-h_i-h_j}\cdot\left(\bp ac{k_\gamma}(y) + \cdots \right) 
     && x>y
     \label{OPE-definitions-last}
\end{align}}
The omissions stand for an infinite sum of descendants of the primary
field in question.
Eqn.~\eqref{OPE-definitions-first}--\eqref{OPE-definitions-last}
define the three sets of structure
constants which are necessary to compute 
the correlation functions of the minimal model under
consideration: the bulk structure constants
$\C{i_\alpha}{j_\beta}{k_\gamma}$ with three bulk fields $i_\alpha$,
$j_\beta$, $k_\gamma$, the bulk-boundary couplings
$\B a{i_\alpha}{k_\gamma}$ with boundary condition $a$, bulk field
$i_\alpha$ and boundary field $k_\gamma$ and the boundary structure
constants $\bc abc{i_\alpha}{j_\beta}{k_\gamma}$ with boundary
conditions $a,b,c$ and boundary fields $i_\alpha$, $j_\beta$,
$k_\gamma$.

By taking different limits in correlation functions one obtains a set
of constraints, called sewing constraints. For orientable genus zero
surfaces it is enough to consider the following four constraints
\cite{Lew92}: 

For four boundary fields ${i_\alpha},{j_\beta},{k_\gamma},{\l_\delta}$
and boundary conditions $a,b,c,d$: 
\begin{align}
      &\sum_{\eps} 
          \bc bcd{j_\beta}{k_\gamma}{q_\eps} 
          \bc abd{i_\alpha}{q_\eps}{\l_\delta} 
          \bc ada{\l_\delta}{\l_\delta}1 \bnpt 1a{UHP} \notag\\
      &\qquad = \sum_p \left( \sum_{\nu} 
          \bc abc{i_\alpha}{j_\beta}{p_\nu} 
          \bc cda{k_\gamma}{\l_\delta}{p_\nu} 
          \bc aca{p_\nu}{p_\nu}1 \bnpt 1a{UHP} \right) \F ijk\l pq
      \label{sewing:bbbb}
\end{align}

For two boundary fields ${p_\nu},{q_\eps}$ and one bulk field ${i_\alpha}$:
\begin{align}
      &\sum_{\delta} 
          \B b{i_\alpha}{\l_\delta} 
          \bc abb{p_\nu}{\l_\delta}{q_\eps} 
          \bc aba{q_\eps}{q_\eps}1 \bnpt 1a{UHP} \notag\\
      &\qquad = \sum_k \left(\sum_{\gamma} 
          \B a{i_\alpha}{k_\gamma} 
          \bc aba{p_\nu}{q_\eps}{k_\gamma} 
          \bc aaa{k_\gamma}{k_\gamma}1 \bnpt 1a{UHP} \right) \notag\\
      &\hspace{2cm}\cdot\sum_m e^{i\pi(2h_m-2h_i-h_p-h_q+\frac 12(h_k+h_\l))}
          \cdot \F i{\ib}qpkm \F pi{\ib}qm\l
      \label{sewing:Bbb}
\end{align}

For two bulk fields ${i_\alpha},{j_\beta}$ and one boundary field ${k_\gamma}$:
\begin{align}
     &\sum_{\rho} 
         \C{i_\alpha}{j_\beta}{m_\rho} 
         \B a{m_\rho}{k_\gamma}
         \bc aaa{k_\gamma}{k_\gamma}1 \bnpt 1a{UHP}\notag\\
     &\qquad = \sum_{p,q} \left(\sum_{\nu,\eps} 
         \B a{i_\alpha}{p_\nu} 
         \B a{j_\beta}{q_\eps} 
         \bc aaa{p_\nu}{q_\eps}{k_\gamma} 
         \bc aaa{k_\gamma}{k_\gamma}1 \bnpt 1a{UHP}\right)\notag\\
     &\qquad\qquad \cdot \sum_{r}
         e^{i\frac\pi{2}(h_k+h_p-h_q-2h_r+h_m-\hb_m-h_i+
             \hb_i+h_j+\hb_j)} \notag\\
     &\qquad\qquad\qquad\cdot
         \F pk{\jb}jqr \F i{\ib}rjpm \F m{\ib}{\jb}kr{\bar m}
     \label{sewing:BBb}
\end{align}

For four bulk fields ${i_\alpha},{j_\beta},{k_\gamma},{\l_\delta}$:
\begin{align}
     &\sum_{\eps} 
         \C{i_\alpha}{k_\gamma}{q_\eps} 
         \C{j_\beta}{\l_\delta}{q_\eps} 
         \C{q_\eps}{q_\eps}1 \npt{1} \notag\\
     &\qquad = \sum_{p,\bar p} \left(\sum_{\nu} 
         \C{i_\alpha}{j_\beta}{p_\nu} 
         \C{k_\gamma}{\l_\delta}{p_\nu} 
         \C{p_\nu}{p_\nu}1  \npt{1} \right)\notag\\
     &\qquad\qquad\cdot e^{i\pi(h_i-\hb_i+h_\l-\hb_\l-h_p+\hb_p-h_q+\hb_q)} \F ij\l kpq
         \F{\ib}{\jb}{\bar \l}{\bar k}{\bar p}{\bar q}
      \label{sewing:BBBB}
\end{align}

These constraints have been rederived using notation and
techniques for calculating with conformal blocks presented in
\cite{MSb90}. Equations \eqref{sewing:bbbb}--\eqref{sewing:BBBB}
require a specific renormalisation of the F--matrices. This
normalisation, together with the explicit derivation of
the constraint~\eqref{sewing:BBb} as an example are given in Appendix~A. 

The form of the constraints given here is slightly different from
\cite{Lew92}. First of all only the Virasoro symmetry was used in
their derivation and not the maximally extended chiral algebra.
Secondly slightly different limits in the correlators were
chosen so that that the expectation value of the identity field
cancels from all expressions. Furthermore the present notation takes
into account fields with multiplicities (see also \cite{BPZ99}).
Leaving out the sums over multiplicities and the multiplicity indices,
\eqref{sewing:bbbb}--\eqref{sewing:BBBB} can be rearranged to match
the corresponding equations in \cite{Lew92}.

\section{Cylinder partition function}
\label{Cylinder-partition-function}

First a note about a convention used throughout this paper. To take
care of the redundancy in the labelling of representations $i$
with Kac-labels $(r,s)$ we will only consider pairs where $r$ is
odd. The set of entries in the Kac-table with $r$ odd is in one-to-one
correspondence to Virasoro highest weight representations of the given
central charge.

Before we can begin to solve \eqref{sewing:bbbb}--\eqref{sewing:BBBB}
we need to know which fields are present in the theory. Since our
starting point is going to be the boundary theory the first thing to
determine is the boundary field content on the upper half plane or,
equivalently, the cylinder partition function.

In Ref.~\cite{BPZ98} a method for the construction of 
the cylinder partition functions associated to a pair of Lie-algebras
$A_n$, $G$ is given. Here we give a quick summary of this method. 

\subsection{General construction}
\label{General-construction}

First choose an odd number $p$ and a number $q$ coprime to $p$.
We will construct the cylinder partition function of
a minimal model with the following central charge:
\begin{align}
  c &= 1 - 6\frac{(p-q)^2}{pq}
\end{align}

Let $A$ be the adjacency matrix of the Dynkin diagram
associated to the Lie-algebra $A_{p-1}$ and $G$ be the adjacency
matrix for a Lie-algebra with Coxeter number $q$. 

For $X{=}A$ or $X{=}G$ define the matrix valued functions $V_n(X)$
recursively via 
\begin{align}
  V_n(X) &= V_2(X)V_{n-1}(X)-V_{n-2}(X) & ;\;\; V_1(X)=\text{id}
  \text{ and } V_2(X)=X
\end{align}
The $V_n$ are called fused adjacency matrices and form a
representation of the Verlinde fusion algebra.

Let $\alpha$ be an odd node of $A$ and $\beta$ be any node of
$G$. Then $a{=}(\alpha,\beta)$ labels the possible boundary
conditions. Let $a{=}(\alpha,\beta)$ and
$b{=}(\bar\alpha,\bar\beta)$. Then the partition function of a cylinder
of circumference $T$ and length $L$ 
with boundary conditions $a$ and $b$ is given by:
\begin{align}
  Z_{a|b} &= \underset{s=1..q-1}{\underset{r=1..p-1}{\sum}}
  {V_r(A)_\alpha}^{\bar\alpha} {V_s(G)_\beta}^{\bar\beta}
  \cdot \chi_{r,s}(q) & ; q=\exp(-\pi T/L)
  \label{construct-cyl-Z}
\end{align}
When we take the identity $\chi_{r,s} = \chi_{p-r,q-s}$ for 
Virasoro characters into account we can rewrite
\eqref{construct-cyl-Z} in the following unique way:
\begin{align}
  Z_{a|b} &= \sum_{i=(r\text{ odd},s)} \n iab \cdot \chi_i(q)
  \label{cylinder-partition-function}
\end{align}

In Refs.~\cite{BPZ98} the genus one sewing constraint for a cylinder
with no field insertions was analysed with the result that $G$ has to be the
adjacency matrix of an A--D--E type Dynkin diagram or of 
a tadpole diagram.

The numbers $\n iab$ can be interpreted as the number of times the
representation $i$ occurs in between the boundary conditions $a$ and
$b$ and they thus describe the field content on the boundary (in this
case the real line).

\subsection{Example: the A-series}
\label{Example:-the-A-series}

For the pair of Lie algebras $(A_{p-1}, A_{q-1})$
one finds $\n iab = \N iab$,
i.e.\ the field content is just given by the Verlinde fusion numbers \cite{CaL91}.
Distinct boundary conditions are given by 
pairings of an odd node in the first A-diagram with any node in the
second:
\begin{align}
  \left\{ \AdiagB{p} , \AdiagC{q} \right\}
\end{align}
There is a distinguished boundary condition, which we will call 1--boundary,
that corresponds to the first node in each $A$-diagram, i.e.\ the $(1,1)$-node.

There are no fields with multiplicity and there is a unique field
$i{=}a$ between the $a$-- and $1$--boundary:
$\bfb ai1 \Rightarrow i{=}a$.
The formula $\n iab = \N iab$ can be understood as saying that the
representations $i$ that can live between the $a$-- and $b$--boundary are exactly 
those occurring in the fusion of the representations $a$ and $b$:
$\bfbfb aa1bb\rightarrow \bfb aib$.

\subsection{The D-series}
\label{The-D-series}

Here $q$ has to be an even number and the boundary conditions are
given by pairings 
of an odd node in the A-diagram with any node in the D-diagram:
\begin{align}
  \left\{ \AdiagB{p} , \DdiagB{q} \right\}
  \label{D-series-boundary-conditions}
\end{align} 
In particular the total number of boundary conditions is
$\frac 14 (p-1)(q+2)$.
Again the boundary condition associated with the first node of each
diagram, i.e.\ the $(1,1)$-node, will get a special name. It will be
called $\o$ so that it is not confused with the $1$--boundary in the
A-series. The $\o$--boundary will play a role similar to that
of the $1$--boundary in \cite{Run99}.

When looking at the D-series boundary field content one observes the
following: 
The boundary conditions can be organised in two categories. In the 
first case, which we will denote as `i-type' boundaries, a boundary
condition $x{=}(\alpha,\beta)$ is associated with an odd node $\alpha$
of the $A$-diagram and any node $\beta$ of the $D$-diagram except for
the two at the split end. The second case we will call `n-type'. An
n-type boundary $a$ is associated with any odd $A$-node and one of the
two end nodes in the $D$-diagram.
\begin{align}
  \text{i-type: }\DdiagC \qquad & \qquad
  \text{n-type: }\DdiagD
\end{align}
The names i-type and n-type stand for `invariant' and `non-invariant'
and are related to the $\Z_2$--symmetry of the D-diagram.

Carrying through the procedure outlined in
section~\ref{General-construction} one finds that 
an n-type boundary $a{=}(\alpha,\beta)$ has exactly one field living on
\bfb a{}\o, denoted by $a_u$ (`unique'). It has representation
labels $(\alpha, \frac q2)$.

An i-type boundary $x{=}(\alpha,\beta)$ has two fields fields living on
\bfb x{}\o with Kac-labels $(\alpha,\beta)$ and
$(\alpha,q-\beta)$ which we will denote by $x_e$ and $x_o$ ($x$ `even'
and $x$ `odd'). The labelling is arbitrary, but we choose to always
give the identity field on the $\o$--boundary an `even' label.

Note that since the Kac-labels of the representations that live
on \bfb x{}\o are given in terms of the node
labels of the boundary condition $x{=}(\alpha,\beta)$ some confusion
might arise. But from the context it will be clear which one of the
two interpretations makes sense.

For definiteness we will fix a specific even/odd labelling for
boundary fields. It will turn out later that the
structure constants involving only i-type boundaries
have an explicit $\Z_2$--symmetry that sends even
fields to themselves and odd fields to minus themselves. This symmetry is
not assumed, but a consequence of the constraints and the gauge we
choose. Following the
arguments in \cite{Rue98} we would like the ground state, i.e.\ the state of
lowest conformal weight, for each pair of boundary conditions to be
invariant under that symmetry.
This can be achieved for all pairs \bfb x{}\o with
$x{=}(\alpha,\beta)$ an i-type boundary by assigning $e/o$-labels
to the two representations living between these boundary conditions
in the following way:
\begin{align}
  x_e = \begin{cases}
    (\alpha,\beta)   & :\; \alpha<\tfrac p2 \\
    (\alpha,q-\beta) & :\; \alpha>\tfrac p2
    \end{cases}
  \qquad & \qquad
  x_o = \begin{cases}
    (\alpha,q-\beta) & :\; \alpha<\tfrac p2 \\
    (\alpha,\beta)   & :\; \alpha>\tfrac p2
    \end{cases}
  \label{even-odd-weights}
\end{align}
As discussed in \cite{Rue98} the physical motivation for this choice
comes from relating the $\Z_2$--action to the effect of a disorder
line stretching from boundary to boundary on a cylinder and trying to
interpret the resulting amplitude as a partition function.
It is also found there that in general it is not possible to fix a
labelling s.t.\ the ground state is invariant for all possible pairs
of boundary conditions. 

The conformal weight of a highest weight representation with
Kac-labels $(r,s)$ is given by
$h_{r,s} = \tfrac{1}{4pq}((qr-ps)^2-(p-q)^2)$.
Thus with definition \eqref{even-odd-weights} we find:
\begin{align}
  h(x_o) - h(x_e) &= \left| (\alpha-\tfrac p2)(\beta-\tfrac q2)\right|
  \label{h-conj-to-h-difference} 
\end{align}

In particular the $\o$--boundary itself is of i-type and it has two
fields living on it, the identity $\o_e{=}1$ and the field $\o_o$ with
$h(\o_o)=(\frac p2{-}1)(\frac q2{-}1)$. We see that $h(\o_o)$ is
integer for $\frac q2$ odd and half-integer for $\frac q2$ even.

We will now proceed to assign e/o/u-labels to all boundary fields, not
just the ones adjacent to the \o--boundary, in the following way: Any
field adjacent to an n-type boundary will get the label $u$. In this
case no multiplicities do occur. In the case $\bfb x{\l_\delta}y$ where both
$x$ and $y$ are of i-type, the possible representations $\l$ are
those that occur in the fusion of $x_e,y_e$ (which is the same set as
in the fusion of $x_o,y_o$) or $x_e,y_o$ (which is the same as
$x_o,y_e$). If $\l$ occurs in the $x_e,y_e$--fusion it gets an
$e$-label, i.e.\ $\delta{=}e$ and if it occurs in the $x_e,y_o$--fusion an
$o$-label $\delta{=}o$. If $\l$ occurs in both fusions, this representation
has multiplicity two and the two corresponding fields have labels
$\l_e$ and $\l_o$.

\begin{table}
\begin{center}
\begin{tabular}{rcll}
\hline
boundary types           & $\l$--labels & field content \\
\hline
i--i  \bfb x\l y & e,o    & $\n\l xy = \N\l{x_e}{y_e} + \N\l{x_e}{y_o}
                                     = \N\l{x_o}{y_o} + \N\l{x_o}{y_e}$ \\
n--i  \bfb a\l y & u      & $\n\l ay = \N\l{a_u}{y_e} = \N\l{a_u}{y_o}$ \\
n--n  \bfb a\l b & u      & $\beta_a\neq \beta_b : \n\l ab = \N\l{a_u}{b_u}$ if
                            $s_\l\equiv 3\mod 4$ \\
                 &        & \phantom{$\beta_a\neq \beta_b :$} and $\n\l ab=0$ otherwise \\
                 &        & $\beta_a=\beta_b : \n\l aa = \N\l{a_u}{a_u}$ if
                            $s_\l\equiv 1\mod 4$ \\
                 &        & \phantom{$\beta_a\neq \beta_b :$} and $\n\l ab=0$ otherwise\\
\hline
\end{tabular}
\end{center}
\caption{Field content and labels between the two types
of boundary conditions.}
\label{tab:bc}
\end{table}

The field content between different boundary conditions is summed up
in table~\ref{tab:bc} listing the numbers $\n iab$ from
\eqref{cylinder-partition-function} for the different cases (here the
representation $\l$ has Kac-labels $\l=(r_\l,s_\l)$ and the boundary
conditions $a,b$ have labels $(\alpha_a,\beta_a)$ and
$(\alpha_b,\beta_b$)).

Notice the double role of the $e/o/u$--indices: When $x$ is a boundary
condition, then $x_e$, $x_o$ or $x_u$ denote the even/odd/unique field
living between the $x$-- and $\o$--boundary. In particular $x_e$ and
$x_o$ denote different representations. If $\l$ is a representation,
then $\l_e$, $\l_o$ or $\l_u$ all denote the same representation and
distinguish fields with multiplicities by labelling one as even and
the other as odd.

\subsection{Ordering of boundary conditions} 
\label{Ordering-of-boundary-conditions}

The constraint equations \eqref{sewing:bbbb}--\eqref{sewing:BBBB}
allow for a large amount of gauge freedom of the structure
constants. Here we find a particular set of structure constants
and show that any solution of the constraints can be transformed into
this set by a regauging. For the construction of the structure constants
presented here an ordering of 
the boundary conditions has to be introduced.
Let $x{=}(\alpha,\beta)$ and $y{=}(\bar\alpha,\bar\beta)$. Then we define:
\begin{align}
  x<y \quad &\Leftrightarrow \quad  (\beta<\bar\beta) \text{ or }
  (\beta=\bar\beta \text{ and } \alpha<\bar\alpha)
\end{align}

\section{Boundary structure constants}
\label{Boundary-structure-constants}

In this section we will construct a set of boundary structure
constants that solve the sewing
constraint \eqref{sewing:bbbb} under the condition that there exists a
solution at all (for the given field content). 
The line of argument in this section is as follows: We assume that
there exists a solution to \eqref{sewing:bbbb} for the D-series field
content. Then we use the property of
\eqref{sewing:bbbb} that a change of basis in the set of primary
boundary fields maps a solution of \eqref{sewing:bbbb} to a new
solution. By redefining the boundary fields one by one this freedom is
exploited to adjust a subset of boundary structure constants to the
values we want. This subset together with \eqref{sewing:bbbb} then
fixes the value of a general boundary structure constant. Altogether
the statement is that any solution can via change of basis be brought to
the form presented in the end of this section.

As input for the boundary solution we use
\begin{itemize}
\item[a)] the $D$-series field content obtained in
  section \ref{The-D-series}
\item[b)] the assumption that all two-point functions 
  $\npt{\bp ab{i_\alpha}(x) \bp ba{i_\alpha}(y)}$ are nonzero, 
  i.e.\ $\bc aba{i_\alpha}{i_\alpha}1{\neq}0$ for all
  $a,b,{i_\alpha}$ that are allowed by a).
\end{itemize}

An interpretation of assumption~b) is
that a zero two-point function implies that the field
in question does not occur in our 
model and hence we are looking at a field content
different from what we demanded in a). This can be seen as follows:
Suppose $\bc aba{i_\alpha}{i_\alpha}1{=}0$ for some choice of
$a,b,{i_\alpha}$. Consider a correlator which contains the field
$\bp ab{i_\alpha}$. Any correlator can be 
expressed as a sum of conformal blocks with coefficients given as
products of structure constants. It is always possible
to take a limit of this correlator where all bulk fields are taken to
the boundary and then all boundary fields except for $\bp ab{i_\alpha}$ are
taken together. In the end we are left with the two-point function 
$\npt{\bp ab{i_\alpha}(x) \bp ba{i_\alpha}(y)}$. Thus in this limit
the coefficient in front of each conformal block contains the factor
$\bc aba{i_\alpha}{i_\alpha}1$ and 
hence the correlator is identically zero. Therefore any
correlator involving $\bp ab{i_\alpha}$ vanishes and this field can as
well be removed from the model, so that we effectively have a field
content different from a).

The actual calculation of the boundary structure constants
is a slightly lengthy case by case study and is presented only for
completeness. The results are summerised in section
\ref{Collection-of-results}.

\subsection{Nonzero boundary structure constants}
\label{Nonzero-boundary-structure-constants}

First we investigate the consequence of b), i.e.\ that all two-point
functions are nonzero. Eqn.~\eqref{sewing:bbbb} can be rewritten so
that the two-point structure constants cancel. To do so we set $\l{=}1$
and obtain the three-point identity:
\begin{align}
  \bfbfbfb a{i_\alpha}b{j_\beta}c{k_\gamma}a \; &\Rightarrow \quad
  \bc bca{j_\beta}{k_\gamma}{i_\alpha} 
  \bc aba{i_\alpha}{i_\alpha}1 = 
  \bc abc{i_\alpha}{j_\beta}{k_\gamma} 
  \bc aca{k_\gamma}{k_\gamma}1
  \label{boundary-3-pt}
\end{align}
Applying this to the r.h.s.\ of \eqref{sewing:bbbb} we get:
\begin{align}
      &\bfbfbfbfb a{i_\alpha}b{j_\beta}c{k_\gamma}d{\l_\delta}a \notag\\
      &\qquad \Rightarrow \quad
      \sum_{\eps} 
      \bc bcd{j_\beta}{k_\gamma}{q_\eps} 
      \bc abd{i_\alpha}{q_\eps}{\l_\delta}
      = \sum_{p,\nu} 
      \bc abc{i_\alpha}{j_\beta}{p_\nu} 
      \bc acd{p_\nu}{k_\gamma}{\l_\delta} \F ijk\l pq
      \label{reduced-bbbb}
\end{align}

Note that there is no freedom the rescale the identity field $1$ on
any boundary. Its normalisation is already fixed by the condition
that $1\cdot\bp xy{i_\alpha}{=}\bp xy{i_\alpha}$, i.e.\ that
$\bc xxy1{i_\alpha}{j_\beta}{=}\delta_{i,j}\delta_{\alpha,\beta}$ and
similar for $\bc xyy{i_\alpha}1{j_\beta}$.

Define an operation `${}^*$' on the
labels $e,o,u$ as follows: 
${x_e}^*{=}x_o$, ${x_o}^*{=}x_e$ and ${x_u}^*{=}x_u$, i.e.\ in terms
of Kac-labels: $(r,s)^* = (r,q-s)$.
Take $\bfbfbfbfb{\o}{\o_o}{\o}{x_i}{x}{x_i}{\o}{\o_o}{\o}$ with
$q{=}1$ and $x$ any boundary condition.
The sum on the r.h.s.\ of \eqref{reduced-bbbb} reduces to
${p_\nu}{=}{x_i}^*$ because the representations $x_i$ and $\o_o$ can fuse
only to ${x_i}^*$. We get:
\begin{align}
      \bc{\o}{x}{\o}{x_i}{x_i}{1} &= 
      \bc{\o}{\o}{x}{\o_o}{x_i}{{x_i}^*}\bc{\o}{x}{\o}{{x_i}^*}{x_i}{\o_o}
      \F{x_i}{\o_o}{\o_o}{x_i}{{x_i}^*}1 \neq 0
      \label{boundary-2pt-nonzero-1}
\end{align}
The l.h.s.\ is nonzero due to assumption b) and hence all terms on the
r.h.s.\ have to be nonzero for a solution with properties a) and b).

In the same way, for an n-type boundary $a$ and an arbitrary boundary
$x$ we can consider 
$\bfbfbfbfb{\o}{x_i}{x}{k_u}{a}{k_u}{x}{x_i}{\o}$, again with
$q{=}1$. The sum reduces to ${p_\nu}{=}a_u$ and we are left with:
\begin{align}
      \bc{x}{a}{x}{k_u}{k_u}{1} &= 
      \bc{\o}{x}{a}{x_i}{k_u}{a_u} \bc{\o}{a}{x}{a_u}{k_u}{x_i}
      \F{k}{x_i}{x_i}{k}{a_u}1 \neq 0
      \label{boundary-2pt-nonzero-2}
\end{align}
Again all terms on the r.h.s.\ have to be nonzero.

\subsection{Mixed boundaries}

Define the boundary condition label $\mu$ as $\mu{=}(1,\frac q2)$,
i.e.\ the first node of the A-diagram and the upper
end node of the D-diagram in
\eqref{D-series-boundary-conditions}. Consider \eqref{reduced-bbbb} with 
$\bfbfbfbfb \mu{\mu_u}\o{\o_o}\o{\o_o}\o{\mu_u}\mu$ and $q_\eps{=}1$. The
sum on the r.h.s.\ reduces to ${p_\nu}{=}\mu_u$ and we are left with:
\begin{align}
  \bc\o\o\o{\o_o}{\o_o}1 = 
  \left(\bc \mu\o\o{\mu_u}{\o_o}{\mu_u}\right)^2 
  \F{\mu_u}{\o_o}{\o_o}{\mu_u}{\mu_u}1
  \label{omega-two-pt-bsc}
\end{align}
Using \eqref{boundary-3-pt} in the form 
$\bfbfbfb{\mu}{\mu_u}{\o}{\o_o}{\o}{\mu_u}{\mu}$
implies 
$\bc{\mu}{\o}{\o}{\mu_u}{\o_o}{\mu_u}{=}\bc{\o}{\o}{\mu}{\o_o}{\mu_u}{\mu_u}$.
We start by rescaling $\bp\o\o{\o_o}$ such that:
\begin{align}
  \bc\o\o\mu{\o_o}{\mu_u}{\mu_u} &= 1
  \label{norm:o-o-mu}
\end{align}
From \eqref{omega-two-pt-bsc} it now follows that
\begin{align}
  \bc{\o}{\o}{\o}{\o_o}{\o_o}{1} &= A & \text{where }
  A = \F{\o_o}{\mu_u}{\mu_u}{\o_o}{\mu_u}{1}
  \label{A-definition}
\end{align}
defining the constant $A$. We use the freedom to rescale
$\bp{\o}{x}{x_o}$ to fix $\bc{\o}{\o}x{\o_o}{x_o}{x_e}{=}A$ for any
i-type boundary $x$.

Let $a$ be an n-type boundary. Taking 
$\bfbfbfbfb \o{\o_o}\o{\mu_u}\mu{{\l_u}}a{a_u}\o$ and ${q_\eps}{=}a_u$ the sum
reduces to ${p_\nu}{=}\mu_u$ and we see, using \eqref{norm:o-o-mu}:
\begin{align}
  \bc\o\mu{a}{\mu_u}{\l_u}{a_u} \bc\o\o a{\o_o}{a_u}{a_u}
  &= \bc\o\mu{a}{\mu_u}{\l_u}{a_u} \F{\mu_u}{\o_o}{a_u}{\l}{\mu_u}{a_u}
  \label{aux-a-eqn}
\end{align}
The representation ${\l_u}$ lives on $\bfb\mu{}{a}$ and the fusion
$\mu_u,a_u\rightarrow{\l_u}$ exists by construction of the field content.
$\bc\o\mu{a}{\mu_u}{\l_u}{a_u}$ is nonzero (see \eqref{boundary-2pt-nonzero-2}) and
the F-matrix element has to be independant of the specific choice of ${\l_u}$ we
make. We denote this F-matrix element with $B_a$:
\begin{align}
  B_a &= \F{\mu_u}{\o_o}{a_u}{\l}{\mu_u}{a_u} &
  \text{where $\bfb\mu{{\l_u}}a$}
  \label{B-definition}
\end{align}
Comparing to \eqref{aux-a-eqn} we see 
$\bc\o\o a{\o_o}{a_u}{a_u}{=}B_a$. It is also useful to define a
constant $C_x$ for any boundary $x$ as 
follows:
\begin{align}
  \text{$x$ of i-type:}\quad
  C_x = \F{x_o}{\o_o}{\o_o}{x_o}{x_e}{1} \quad&\quad
  \text{$x$ of n-type:}\quad
  C_x = \F{x_u}{\o_o}{\o_o}{x_u}{x_u}{1}
  \label{C-definition}
\end{align}
From \eqref{reduced-bbbb} with 
$\bfbfbfbfb \o{\o_o}\o{\o_o}\o{a_u}a{a_u}\o$, $q_\eps{=}a_u$ and
$p_\nu{=}1$ we get the following identity for any n-type boundary $a$:
\begin{align}
   C_a \cdot B_a = \frac{A}{B_a} \label{ABC-relation} 
\end{align}

Let $x$ and $y$ be two boundaries s.t.\ one is of n-type. Then by
rescaling $\bp xy{\l_u}$ for $x{\neq}\o$ and $\l{\neq}1$ we fix
$\bc{\o}{x}{y}{x_\alpha}{\l_u}{y_\beta}{=}1$ where $\alpha$, $\beta$
stand for the labels $e$ or $u$, as applicable for the present
boundary conditions.  
Let $x$ be an i-type boundary. We can use \eqref{reduced-bbbb} to
calculate the structure constants involving $x_o$ instead of $x_e$:
\begin{align}
  \bc{\o}{x}{a}{x_o}{\l_u}{a_u} = 
  \frac{A}{B_a} \F{x_o}{\o_o}{a_u}{\l}{x_e}{a_u} &&
  \bc{\o}{a}{x}{a_u}{\l_u}{x_o} =
  \frac{B_a}{A} \F{a_u}{\o_o}{x_e}{\l}{a_u}{x_o}
  \label{initial-wax-form-xo}
\end{align}

Using \eqref{reduced-bbbb} and the identities in appendix~B 
one can verify that the definitions above give rise to
symmetrical structure constants, i.e.\ 
$\bc{\o}{x}{y}{x_\alpha}{\l_u}{y_\gamma}
{=}\bc{y}{x}{\o}{\l_u}{x_\alpha}{y_\gamma}$ 
where at least one of $x,y$ is of n-type.

\subsection{i-type boundaries}
\label{i-type-boundaries}

Let $x,y$ be i-type boundaries. Consider the structure constant 
$\bc{\o}xy{x_\alpha}{\l_\beta}{y_\gamma}$ where the labels
$\alpha,\beta,\gamma$ are either $e$ or 
$o$ (as $u$ cannot occur in this situation). We will now try to
regauge the boundary fields such that the following rule holds:

\begin{itemize}
\item[] The structure constant
$\bc{\o}xy{x_\alpha}{\l_\beta}{y_\gamma}$ can be 
nonzero only if $\{\alpha,\beta,\gamma\}$ is one of the (unordered)
sets $\{e,e,e\}$ or $\{e,o,o\}$. 
\end{itemize}

Suppose that the representation $\l$ on $\bfb x{}y$ occurs with
multiplicity one. Recall from the analysis of the D-series boundary
field content in section \ref{The-D-series} that then
the above rule is automatically true, since
e.g.\ if $\l$ gets label $e$ then it occurs in the fusion of $x_e,y_e$
or $x_o,y_o$ but not in the fusion of  $x_e,y_o$ and $x_o,y_e$.

If the representation $\l$ occurs with multiplicity two, we can use
the freedom to form linear combinations of the two fields to make the
above rule true. This is explained in more detail below.

Let $\l$ be a representation that occurs with multiplicity two. Let 
$\bp xy{\l_e}$ and $\bp xy{\l_o}$ be the two fields. Recall that we
normalised the two-point functions in such a way that only 
$\npt{\bp xy{\l_e}\bp yx{\l_e}}$ and $\npt{\bp xy{\l_o}\bp yx{\l_o}}$ are
nonzero. When taking linear combinations of the primary fields one
has to preserve this condition, or the form of the sewing constraints
would change. One gets the following constraints:
\begin{align}
  \npt{\bp xy{\l_e}\bp yx{\l_o}} = 0 \qquad , & \qquad 
  \npt{\bp xy{\l_o}\bp yx{\l_e}} = 0
  \label{bnd-2pt-condition}
\end{align}
Now consider the following change of basis:
\begin{align}
  {\begin{pmatrix} \bp xy{\l_e} \\ \bp xy{\l_o} \end{pmatrix}}_{\!\!\text{new}}
  =\;
  \begin{pmatrix} a & b \\ c & d \end{pmatrix}
  {\begin{pmatrix} \bp xy{\l_e} \\ \bp xy{\l_o} \end{pmatrix}}_{\!\!\text{old}}
  \quad , & \quad
  {\begin{pmatrix} \bp yx{\l_e} \\ \bp yx{\l_o} \end{pmatrix}}_{\!\!\text{new}}
  =\;
  \begin{pmatrix} \tilde a & \tilde b \\ \tilde c & \tilde d \end{pmatrix}
  {\begin{pmatrix} \bp yx{\l_e} \\ \bp yx{\l_o} \end{pmatrix}}_{\!\!\text{old}}
\end{align}
where the two matrices are invertible. The constraints
\eqref{bnd-2pt-condition} amount to:
\begin{align} 
  a \tilde c \cdot \bc xyx{\l_e}{\l_e,{\text{old}}}1 + b \tilde d 
  \cdot \bc xyx{\l_o}{\l_o,{\text{old}}}1 = 0 \quad&\quad
  c \tilde a \cdot \bc xyx{\l_e}{\l_e,{\text{old}}}1 + d \tilde b 
  \cdot \bc xyx{\l_o}{\l_o,{\text{old}}}1 = 0 
\end{align}
We see that we can choose the new basis of $\bp xy{}$ arbitrarily and
the above condition fixes the direction, but not the length of the new
basis vectors for $\bp yx{}$.

This change of basis can be used to bring the following $2{\times}2$--matrix
composed of structure constants to diagonal form:
\begin{align}
  \begin{pmatrix}
    \bc\o xy{x_e}{\l_e}{y_e} & \bc\o xy{x_e}{\l_e}{y_o} \\
    \bc\o xy{x_e}{\l_o}{y_e} & \bc\o xy{x_e}{\l_o}{y_o}
  \end{pmatrix}
  \quad \longrightarrow \quad
  \begin{pmatrix} * & 0 \\ 0 & * \end{pmatrix}
\end{align}
We now redefined the structure constants
$\bc\o xy{x_\alpha}{\l_\beta}{y_\gamma}$ in such a way that they obey
our desired coupling relations $\{e,e,e\}$ or $\{e,o,o\}$. 

The change of basis fixes the new $\bp yx{\l_e}$ and 
$\bp yx{\l_o}$ up to rescaling. However they do already obey the
odd/even-coupling relation. To see this consider \eqref{reduced-bbbb}
in the form: 
\begin{align}
      &\bfbfbfbfb \o{x_e}x{\l_e}y{\l_o}x{x_e}\o, q_\eps{=}1 \notag\\
      &\qquad \Rightarrow \quad \bc xyx{\l_e}{\l_o}1
      = \bc\o{x}{y}{x_e}{\l_e}{y_e} \bc\o{y}{x}{y_e}{\l_o}{x_e} \F {x_e}{\l}{\l}{x_e}{y_e}1
      \label{works-for-yx-as-well}
\end{align}
The l.h.s.\ of this equation is zero and on the r.h.s.\ both 
$\bc\o{x}{y}{x_e}{\l_e}{y_e}$ and the F-matrix entry are nonzero (this
can be seen from evaluating $\bc xyx{\l_e}{\l_e}1{\neq}0$ instead of 
$\bc xyx{\l_e}{\l_o}1{=}0$ in \eqref{works-for-yx-as-well}). Hence
$\bc\o{y}{x}{y_e}{k_o}{x_e}{=}0$ as we said.

From $\bc{\o}{\o}x{\o_o}{x_o}{x_e}{=}A$ in the previous section
we can determine $\bc{\o}{\o}x{\o_o}{x_e}{x_o}$ by using
\eqref{reduced-bbbb} and \eqref{F-identity-a}:
\begin{align}
  &\bfbfbfbfb{\o}{\o_o}{\o}{\o_o}{\o}{x_e}{x}{x_e}{\o}, q_\eps{=}x_o \notag\\
  &\qquad\Rightarrow\quad \bc{\o}{\o}x{\o_o}{x_e}{x_o} = 
  \frac{\bc{\o}{\o}{\o}{\o_o}{\o_o}{1}}{\bc{\o}{\o}x{\o_o}{x_o}{x_e}}
  \cdot \F{\o_o}{\o_o}{x_e}{x_e}{1}{x_o} = \frac{1}{C_x}
  \label{initial-wwx-form-xo}
\end{align}

Next we fix the form of $\bc{\o}xy{x_\alpha}{\l_\beta}{y_\gamma}$.
For $\l{\neq}1$, $x{\neq}\o$ we rescale $\bp xy{\l_e}$ and
$\bp xy{\l_o}$ s.t.: 
\begin{align}
  \bc{\o}{x}{y}{x_o}{\l_e}{y_o} = \F{x_o}{\o_o}{y_e}{\l}{x_e}{y_o}
  \qquad&\qquad 
  \bc{\o}{x}{y}{x_e}{\l_o}{y_o} = 
  \begin{cases}
    1 & \text{ if $x\leq y$} \\
    \F{x_e}{\o_o}{y_e}{\l}{x_o}{y_o} & \text{ if $x>y$} 
  \end{cases}
  \label{initial-wxy-form}
\end{align}
One can verify that for $\l{=}1$ indeed
$\bc{\o}{x}{x}{x_o}{1}{x_o}{=}1$ and that for $x{=}\o$ the above
reduces to the expressions for $\bc\o\o{x}{}{}{}$ given before.

As in the case of mixed boundaries one can now check that the
definitions in \eqref{initial-wxy-form} give rise to
reflection symmetric structure constants, but this time there is an
exception for $x{=}y$: In this case  
some of the structure constants are
antisymmetric under reflection. As opposed to the A-series, for the D-series it is
in general impossible to find a gauge in which all boundary structure constants
have reflection symmetry. To see this consider \eqref{reduced-bbbb} in
the form:
\begin{align}
      &\bfbfbfbfb {\o}{x_e}{x}{\l_o}{x}{x_e}{\o}{\o_o}{\o} \notag\\
      &\qquad\Rightarrow\quad
      \bc{x}{x}{\o}{\l_o}{x_e}{x_o} \bc{\o}{x}{\o}{x_e}{x_o}{\o_o} =
      \bc{\o}{x}{x}{x_e}{\l_o}{x_o} \bc{\o}{x}{\o}{x_o}{x_e}{\o_o} 
      \F{\l}{x_e}{\o_o}{x_e}{x_o}{x_o}
      \label{bsc-not-symmetric}
\end{align}
Eqn.~\eqref{F-identity-b} forces the F-matrix entry in \eqref{bsc-not-symmetric}
to square to one. But in general it does take the values $\pm 1$, depending on
$\l_o$. So no matter how we choose $\bc{\o}{x}{\o}{x_e}{x_o}{\o_o}$ and 
$\bc{\o}{x}{\o}{x_o}{x_e}{\o_o}$, in general we cannot avoid that if
$\bc{\o}{x}{x}{x_e}{\l_o}{x_o}$ is symmetric for some values of
$\l_o$, it will be antisymmetric for others. 

One can verify that \eqref{initial-wxy-form} implies that for even fields alone 
we get a solution that resembles the A-series, 
i.e.\ $\bc{\o}{x}{y}{x_e}{\l_e}{y_e}{=}1$, whereas the various structure constants
involving odd fields are:
\begin{align}
  \text{For $x\leq y$: } \quad &
    \bc{\o}{x}{y}{x_e}{\l_o}{y_o} = 1 \;;\quad
    \bc{\o}{x}{y}{x_o}{\l_e}{y_o} = \F{x_o}{\o_o}{y_e}{\l}{x_e}{y_o} \;;\notag\\
    & \bc{\o}{x}{y}{x_o}{\l_o}{y_e} = A\cdot\F{x_o}{\o_o}{y_o}{\l}{x_e}{y_e}\cdot C_y \notag\\
  \text{For $x>y$: } \quad &
    \bc{\o}{x}{y}{x_e}{\l_o}{y_o} = \F{x_e}{\o_o}{y_e}{\l}{x_o}{y_o} \;;\quad
    \bc{\o}{x}{y}{x_o}{\l_e}{y_o} = \F{x_o}{\o_o}{y_e}{\l}{x_e}{y_o} \;;\notag\\
    & \bc{\o}{x}{y}{x_o}{\l_o}{y_e} = A\cdot C_x 
    \label{odd-results-i-type}
\end{align}

\subsection{General boundary structure constants}

Consider \eqref{reduced-bbbb} in the form
\begin{align}
      &\bfbfbfbfb {\o}{x_\alpha}x{i_r}y{j_s}z{z_\gamma}{\o}, q_\eps{=}k_t \notag\\
      &\qquad \Rightarrow \quad
      \bc{x}{y}{z}{i_r}{j_s}{k_t} = \sum_\beta
      \frac{\bc{\o}{x}{y}{x_\alpha}{i_r}{y_\beta} 
            \bc{\o}{y}{z}{y_\beta}{j_s}{z_\gamma}}{\bc{\o}{x}{z}{x_\alpha}{k_t}{z_\gamma}}
      \F{i}{x_\alpha}{z_\gamma}{j}{y_\beta}{k}
      \label{general-c-sum}
\end{align}
where the set $\{\alpha,t,\gamma\}$ has to be one of $\{e,e,e\}$,
$\{e,o,o\}$ or $\{u,*,*\}$.
Since all constants that can occur on the r.h.s.\ have been computed
in the previous two sections, the general boundary
structure constants can be obtained from \eqref{general-c-sum}.

Suppose all boundaries are of i-type. Then choosing $\alpha{=}e$ and
$\gamma{=}t$ reduces the sum in \eqref{general-c-sum} to
$\beta{=}r$.In the numerator on the r.h.s.\ we now see the boundary
structure constant $\bc{\o}{y}{z}{y_r}{j_s}{z_t}$, which obeys the
even/odd coupling rule as seen in 
section~\ref{i-type-boundaries}. This implies that 
the even/odd coupling rule extends to all boundary structure
constants, i.e.\ $\bc{x}{y}{z}{i_r}{j_s}{k_t}$ can be nonzero only if 
$\{r,s,t\}$ is one of the sets $\{e,e,e\}$,
$\{e,o,o\}$ or $\{u,*,*\}$.

We can also investigate the behaviour of the boundary structure constants
under reflection, i.e.\ given $\bc{x}{y}{z}{i_r}{j_s}{k_t}$, what is
$\bc{z}{y}{x}{j_s}{i_r}{k_t}$? Using \eqref{boundary-3-pt} we check
that for $\bc{\o}{x}{y}{x_\alpha}{\l_\beta}{y_\gamma}$ reflection symmetry is
equivalent to $\eps{=}1$ in the following equation:
\begin{align}
  \bc{\o}{x}{y}{x_\alpha}{\l_\beta}{y_\gamma} &= \eps \cdot
  \frac{\bc{\o}{x}{\o}{x_\alpha}{x_\alpha}{1}}{\bc{\o}{y}{\o}{y_\gamma}{y_\gamma}{1}}
  \bc{\o}{y}{x}{y_\gamma}{\l_\beta}{x_\alpha}
  \label{alternate-wxy-symmetry}
\end{align}
As remarked in the previous two sections, using the explicit form of
the structure constants derived
there, we see that $\eps{=}1$ in all cases except for one, and that is
$\beta{=}o$ and $x{=}y$, where we can have $\eps{=}\pm 1$. This can be made more
precise in the light of \eqref{bsc-not-symmetric}, if we define:
\begin{align}
  \eps(x,x,\l_o) = \F{\l}{x_e}{\o_o}{x_e}{x_o}{x_o} 
  \quad \text{ and } \quad
  \eps(x,y,\l_\beta) = 1 \text{ in all other cases}
\end{align}
Then \eqref{alternate-wxy-symmetry} holds with $\eps=\eps(x,y,\l_i)$
for all $x,y,\l_i$ and we can use this relation to evaluate the
reflection property of the general boundary structure constants
obtained in \eqref{general-c-sum}. We get:
\begin{align}
  \bc{x}{y}{z}{i_r}{j_s}{k_t} = 
  \frac{\eps(x,y,i_r)\eps(y,z,j_s)}{\eps(x,z,k_t)} \cdot
  \bc{z}{y}{x}{j_s}{i_r}{k_t}
  \label{general-c-sym}
\end{align}

\subsection{Collection of results}
\label{Collection-of-results}

We found: Any solution to the
boundary four-point constraint \eqref{sewing:bbbb} for a given
D-series field content can be brought to the same form by regauging the
fields, i.e.\ if there is a solution, it is essentially unique. 

In the gauge we chose, there
are two conditions that have to be satisfied before a given fusion 
$\bp xy{i_\alpha} \bp yz{j_\beta} \rightarrow \bp xz{k_\gamma}$ can
exist. First the according Verlinde fusion number $\N ijk$ has to be
nonzero and second the $e/u/o$-labels $\{\alpha,\beta,\gamma\}$ have
to be one of the sets $\{u,*,*\}$, $\{e,e,e\}$ or $\{e,o,o\}$ (where
$*$ stands for any label). The assignment of the $e/o/u$-labels is
described in section~\ref{The-D-series}. From this 
even/odd--coupling rule we see immediately that the
boundary structure constants involving only i-type boundaries
have a $\Z_2$--symmetry of the form 
$\bp xy{\l_e}\rightarrow \bp xy{\l_e}$ and 
$\bp xy{\l_o}\rightarrow -\bp xy{\l_o}$.

There are two boundary conditions which we have endowed with a special
name. $\o$ stands for the boundary condition associated to the $(1,1)$
pair of nodes in the diagram \eqref{D-series-boundary-conditions} and
$\mu$ for $(1,\frac q2)$. The boundary structure constants take the
following form:
\begin{enumerate}
\item[$\bullet$] Two-point functions with $\o$ (the constants $A$ and $C_x$ are
  defined in \eqref{A-definition} and \eqref{C-definition}):
  \begin{align}
    \bc{\o}{x}{\o}{x_u}{x_u}{1}=1 \qquad 
    \bc{\o}{x}{\o}{x_e}{x_e}{1}=1 \qquad 
    \bc{\o}{x}{\o}{x_o}{x_o}{1}=A\cdot C_x
    \label{collection-first-bsc-formula}
  \end{align}
\item[$\bullet$] For $x$ any type and $y$ of n-type ($B_y$ is
  defined in \eqref{B-definition}):
  \begin{align}
    \bc{\o}{x}{y}{x_u}{\l_u}{y_u} = 1 \qquad
    \bc{\o}{x}{y}{x_e}{\l_u}{y_u} = 1 \qquad
    \bc{\o}{x}{y}{x_o}{\l_u}{y_u} = 
    \frac{A}{B_y} \F{x_o}{\o_o}{y_u}{\l}{x_e}{y_u} 
  \end{align}
\item[$\bullet$] For $x,y$ of i-type and $x{\leq}y$ (the ordering is defined in 
  section~\ref{Ordering-of-boundary-conditions}):
  \begin{align}
    \bc{\o}{x}{y}{x_e}{\l_e}{y_e} &= 1 &
    \bc{\o}{x}{y}{x_e}{\l_o}{y_o} &= 1 \notag\\
    \bc{\o}{x}{y}{x_o}{\l_e}{y_o} &= \F{x_o}{\o_o}{y_e}{\l}{x_e}{y_o} &
    \bc{\o}{x}{y}{x_o}{\l_o}{y_e} &= A\cdot C_y\cdot\F{x_o}{\o_o}{y_o}{\l}{x_e}{y_e}
  \end{align}
\end{enumerate}
All other cases can be obtained from this list by using
\eqref{alternate-wxy-symmetry} with $\eps{=}1$ (since the case $x{=}y$
is covered in the list), with results listed in
\eqref{initial-wax-form-xo}, \eqref{odd-results-i-type}. 

For the general boundary structure constants we distinguish two
cases. First, for $x,z$ of n-type and $y$ of i-type the sum over
$\beta$ in \eqref{general-c-sum} has to be carried out and we
get:
\begin{align}
      \bc{x}{y}{z}{i_u}{j_u}{k_u} &= 
      \F{i}{x_u}{z_u}{j}{y_e}{k} +\frac{B_x}{B_z}\cdot
      \F{i}{y_e}{\o_o}{x_u}{x_u}{y_o}
      \F{j}{z_u}{\o_o}{y_o}{y_e}{z_u}
      \F{i}{x_u}{z_u}{j}{y_o}{k}
      \label{bsc-sum-not-reduced}
\end{align}
In all other cases the sum reduces to one term with the result:
\begin{align}
      \bc{x}{y}{z}{i_r}{j_s}{k_t} &= 
      \frac{\bc{\o}{x}{y}{x_\alpha}{i_r}{y_\beta} 
            \bc{\o}{y}{z}{y_\beta}{j_s}{z_\gamma}}{\bc{\o}{x}{z}{x_\alpha}{k_t}{z_\gamma}}
      \F{i}{x_\alpha}{z_\gamma}{j}{y_\beta}{k}
      \label{collection-last-bsc-formula}
\end{align}
The r.h.s.\ does not depend on the specific choice of
$\alpha,\beta,\gamma$ as long as the combinations
$\{\alpha,r,\beta\}$, $\{\beta,s,\gamma\}$ and $\{\alpha,t,\gamma\}$ are
allowed by the even/odd coupling rule. The structure constant in the
denominator is then automatically nonzero
(see section \ref{Nonzero-boundary-structure-constants}).
The boundary structure constants are either symmetric or
anti--symmetric under reflection, the precise behaviour is
given in eqn.~\eqref{general-c-sym}.

Note in particular that if $x,y,z$ are all of n-type or if the fields
$i_r$, $j_s$, $k_t$ are even, the solution takes the same form as in
the A-series (see eqn.~(31) in \cite{Run99}):
\begin{align}
      \bc{x}{y}{z}{i_e}{j_e}{k_e} =  \F{i}{x_e}{z_e}{j}{y_e}{k}
      \quad&\text{ and }\quad
      \bc{x}{y}{z}{i_u}{j_u}{k_u} =  \F{i}{x_u}{z_u}{j}{y_u}{k}
      \label{bsc-only-even-fields}
\end{align}
In fact this form also holds for all mixed cases with only $e/u$ field
labels, except when the boundaries $x,z$ are of n-type and $y$ is of
i-type, when we obtained \eqref{bsc-sum-not-reduced}.

\section{Extension to bulk theory}
\label{Extension-to-bulk-theory}

The bulk-boundary couplings $\B a{i_\alpha}{k_\gamma}$ can be
determined by analysing the sewing constraint \eqref{sewing:Bbb} in a
rewritten form using the assumption that all $\bc
xyx{i_\alpha}{i_\alpha}1$ are nonzero for fields present in the model
(see section~\ref{Boundary-structure-constants}): 
\begin{align}
  \sum_{\delta} 
      \B b{i_\alpha}{\l_\delta} 
      \bc abb{p_\nu}{\l_\delta}{q_\eps}
  &= \sum_{k,\gamma}  
      \B a{i_\alpha}{k_\gamma} 
      \bc aab{k_\gamma}{p_\nu}{q_\eps} \cdot\sum_m
          e^{i\pi(2h_m-2h_i-h_p-h_q+\frac 12(h_k+h_\l))} \notag\\
  & \hspace{3cm}\cdot \F i{\ib}qpkm \F pi{\ib}qm\l
  \label{reduced-Bbb}
\end{align}

Suppose now there is a boundary condition $a$ and a bulk field
${i_\alpha}$ s.t.\ ${i_\alpha}$ does not couple to any
boundary field on the boundary $a$,
i.e.\ $\B a{i_\alpha}{k_\gamma}{=}0$ for all ${k_\gamma}$ with
$\n kaa\neq0$. Then, roughly 
speaking (this is
discussed in more detail below), since the
r.h.s.\ of \eqref{reduced-Bbb} is now zero, all $B$'s involving the
bulk field ${i_\alpha}$ are zero.
Since any n-point function has a limit in which we take all bulk fields to
the boundary, and hence in this limit the factors in front of the
conformal blocks 
contain the product of the $B$'s of all bulk fields, any n-point
function involving the field ${i_\alpha}$ is identically zero. This in turn
just means that the bulk field ${i_\alpha}$ is not part of the model
and can be left out.

Using this idea we can now determine the maximal bulk field content
consistent with \eqref{reduced-Bbb}. This does not prove that the
maximal field content has to be modular invariant, but it turns out
that it is indeed the D-series bulk field content derived in
\cite{CIZ87}.

To make this more precise,
the classification of boundary conditions and their field content in
\cite{BPZ98} establishes a one--to--one correspondence between boundary
conditions and diagonal bulk fields. In this approach specifying the
possible boundary conditions is equivalent to giving the diagonal part
of the bulk partition function, and from \cite{CIZ87} one knows which
off-diagonal parts have to be added to make it modular invariant. The
statement thus is that, at least of the A-- and D--series, the
bulk field content obtained by this procedure is the maximal
consistent one.

\subsection{Bulk field content}
\label{Bulk-field-content}

For an arbitrary boundary $x$ consider \eqref{reduced-Bbb} with
$a{=}\o$, $b{=}x$, ${p_\nu}{=}x_\nu$, ${q_\eps}{=}x_\eps$. The ${\l_\delta}$
sum reduces to one element and we get:
\begin{align}
  \B x{i_\alpha}{\l_\delta} \bc{\o}{x}{x}{x_\nu}{\l_\delta}{x_\eps}
  &= \B{\o}{i_\alpha}{1} \bc{\o}{\o}{x}{1}{x_\nu}{x_\eps} \cdot
     \big(\text{ F's }\big)\Big|_{k=1} \notag\\
  &\qquad + 
     \B{\o}{i_\alpha}{\o_o} \bc{\o}{\o}{x}{\o_o}{x_\nu}{x_\eps} \cdot
     \big(\text{ F's }\big)\Big|_{k=\o_o}
  \label{general-xw-Bbb}
\end{align}

First note that from an argument similar to
\eqref{boundary-2pt-nonzero-1} and \eqref{boundary-2pt-nonzero-2} we
see that all boundary structure constants that appear in
\eqref{general-xw-Bbb} are nonzero if they are allowed by fusion and
the even/odd coupling rule (see section
\ref{Nonzero-boundary-structure-constants}). 

Suppose a bulk field ${i_\alpha}$ does not couple to the $\o$--boundary,
i.e.\ $\B{\o}{i_\alpha}{1}{=}0$ and $\B{\o}{i_\alpha}{\o_o}{=}0$. Then the 
r.h.s.\ of \eqref{general-xw-Bbb} is
identically zero and since the boundary structure constant on the l.h.s.\ is nonzero
it follows that $\B a{i_\alpha}{k_\gamma}{=}0$ for all $a$,${k_\gamma}$. As we argued before
this implies that all correlators involving ${i_\alpha}$ vanish and the field
can be removed from the bulk theory. Hence for any bulk field ${i_\alpha}$ we
need at least one of $\B{\o}{i_\alpha}{1}$ and $\B{\o}{i_\alpha}{\o_o}$ to be
nonzero. 

The bulk field ${i_\alpha}$ transforms in a tensor product of two
representations $i{\otimes}\ib$. Suppose
${i_\alpha}$ is a field with spin, i.e.\ $i{\neq}\ib$. Then $i$ and $\ib$
cannot fuse to the identity and hence it follows that $\B{\o}{i_\alpha}1{=}0$. 
So for the field ${i_\alpha}$ to exist we need $\B{\o}{i_\alpha}{\o_o}{\neq}0$. This is only
possible if the fusion $i,\ib\rightarrow\o_o$ exists, i.e.\ only pairs
with Kac-labels $i{=}(r,s)$ and $\ib{=}(r,q-s)$ are possible. Now
consider \eqref{general-xw-Bbb} with $x{=}\o$, $\l_\delta{=}x_\nu{=}\o_o$,
$x_\eps{=}1$. We obtain (compare also to \eqref{h-conj-to-h-difference}):
\begin{align}
  \B{\o}{i_\alpha}{\o_o} &= \B{\o}{i_\alpha}{\o_o} \cdot e^{2\pi i(\hb_i-h_i)}
  =  \B{\o}{i_\alpha}{\o_o} \cdot e^{2\pi i(\frac p2 - r)(\frac q2-s)}
\end{align}
For the exponential to be $+1$ we need
$\frac q2 - s$ to be even (recall
that for the D-series $p$ is always odd and $q$ always even).

Next suppose that the bulk field ${i_\alpha}$ is diagonal, 
i.e.\ $i{=}\ib$. Let $i$ have
Kac-labels $(r,s)$. As before denote by $i^*$ the representation with
Kac-labels $(r,q-s)$. Consider \eqref{general-xw-Bbb} with $x{=}\o$, $\l_\delta{=}1$,
$x_\nu{=}x_\eps{=}\o_o$. The F-sum in \eqref{reduced-Bbb} reduces to $m{=}i^*$
and we obtain:
\begin{align}
  \B{\o}{i_\alpha}{1} &= \B{\o}{i_\alpha}{1} \cdot e^{2\pi i({h_i}^*-h_i-h_{\o_o})}
  =  - \B{\o}{i_\alpha}{1} \cdot e^{-i\pi s}
\end{align}
Hence only diagonal bulk fields with representation
$i{=}(r,s)$ where $s$ is odd are present. Recall that by convention
$r$ is always odd.

Up to now we have determined which pairs of representations are
possible in the bulk. Now we have to determine their
multiplicities. The maximal multiplicity a bulk field can have is
equal to the number of boundary fields it couples to on the
$\o$--boundary. This can be seen as follows: Suppose two bulk fields
$i_\alpha,i_\beta$ in the same representation $i{\otimes}\ib$ can due to the fusion
rules only couple to one field $\o_\gamma$ on the
$\o$--boundary. Then we can define two new bulk fields as linear
combination of ${i_\alpha},{i_\beta}$ s.t., say,
${\B\o{i_\alpha}{\o_\gamma}}_{\hspace{-10pt}\text{,new}}{\neq}0$ and 
${\B\o{i_\beta}{\o_\gamma}}_{\hspace{-10pt}\text{,new}}{=}0$. 
Now the new $i_\beta$ field does not 
couple to the $\o$--boundary at all and can be removed from the theory.
A similar argument forbids more than two bulk fields in the same
representation that by fusion can couple to the two fields on the
$\o$--boundary. The only bulk-representations that allow coupling to
both fields on the $\o$--boundary is the pair $i{\otimes}i$ where
$i$ has Kac-labels $(r,\frac q2)$ and only in this
case multiplicity two can occur.

\begin{table}[t]
\begin{center}
\begin{tabular}{cll}
\hline
$i,\ib$ fuse to & representations \\
\hline
$1$    & $(i,\ib)$ with $i=\ib=(r,s)$ where $r$ and $s$ odd \\
$\o_o$ & $(i,\ib)$ with $i=(r,s)$, $\ib=(r,q{-}s)$ where $r$
               odd and $s{-}\frac q2$ even\\[2pt]
\hline
\\[-25pt]
\end{tabular}
\end{center}
\caption{Maximal bulk field content for D-series}
\label{tab:fc}
\end{table}

Table~\ref{tab:fc} summarizes the maximal bulk field content
consistent with a $D$-type cylinder partition function.
Each representation in the table occurs once for each possible fusion.
The only situation in which multiplicity arises is for pairs of
representations that can fuse to both $1$ and $\o_o$, in which case we
have multiplicity two. Note that as mentioned before this can only
happen for a diagonal bulk field with both representations of the form
$(r,\frac q2)$ and when $\frac q2$ is odd, i.e.\ in $D_{\text{even}}$
models. 

Note that the same line of argument also works for the A-series. Here
the situation is much simpler, since there is only one field, i.e.\ the
identity, on the $1$--boundary. Hence only diagonal bulk fields are
allowed and only multiplicity one can occur. This is precisely the
A-series modular invariant bulk field content.

\subsection{Bulk-boundary couplings}

As a first step to determine the bulk-boundary couplings we will
choose a basis of primary bulk fields s.t.\ each field couples either
to 1 or to $\o_o$ on the $\o$--boundary, but 
not to both simultaneously.

As we have seen in the previous section, this demand is automatically
enforced by the fusion rules except for diagonal bulk fields in the
representation $i{\otimes}i$ where $i$ has Kac-labels $(r,\frac q2)$ and we
are looking at a $D_{\text{even}}$ model. In this case the corresponding
bulk fields have multiplicity two and similar to the argument for
redefining the boundary structure constants in section
\ref{i-type-boundaries} we can take linear combinations of the two
bulk fields to obtain the desired coupling rule. 
We will denote bulk fields that couple to $1$ on the $\o$--boundary as
`even' and bulk fields that couple to the boundary field $\o_o$ as
`odd'. The new bulk--boundary
couplings then fulfill $\B{\o}{i_e}{\o_o}{=}0$ for the even bulk fields
$i_e$ and $\B{\o}{i_o}{1}{=}0$ for the odd fields $i_o$.

We are still free to rescale all the bulk fields ${i_\alpha}$. We use this
freedom to set (the odd diagonal fields are normalised differently so
that the $\B x{i_\alpha}1$ represent the Pasquier algebra as seen in
section~\ref{Disc-partition-function-and-Pasquier-Algebra}):
{\allowdisplaybreaks
\begin{align}
    &\B{\o}{i_o}{\o_o}=\frac{B_i}A \cdot \frac{\S 1i}{\S 11}&&
    \text{for diagonal odd fields $i_o$ (i.e.\ $i{=}\ib$)} \notag\\
    &\B{\o}{i_e}{1}=\B{\o}{i_o}{\o_o}=\frac{\S 1i}{\S 11} 
    \cdot e^{i\frac{\pi}{2}(h_i-\hb_i)} &&
    \text{for all other cases} \notag \\
    &\B{\o}{i_o}1 = \B{\o}{i_e}{\o_o} = 0 &&
    \text{from the coupling rule}  
    \label{bulk-field-normalisation}
\end{align}}

With this redefinition of the fields the sum on the r.h.s.\ of
\eqref{general-xw-Bbb} reduces to one term. Using
\eqref{general-xw-Bbb} with $\eps{=}\delta$ and
$k_\gamma{=}\o_\alpha$ the general bulk--boundary coupling takes the form:
\begin{align}
  \B x{i_\alpha}{\l_\delta} &= \B{\o}{i_\alpha}{\o_\alpha} \cdot \sum_m
     \exp\left(i\pi(2h_m-2h_i-h_{x_\nu}-h_{x_\delta}+\tfrac 12(h_{\o_\alpha}+h_\l))\right)\notag\\
     &\hspace{2cm}\cdot
     \F i{\ib}{x_\delta}{x_\nu}{\o_\alpha}m \F {x_\nu}i{\ib}{x_\delta}m\l
     \cdot
     \begin{cases}
       1          &\text{: $\alpha{=}e$ and $\delta{=}e\text{ or }u$}\\
       {C_x}^{-1} &\text{: $\alpha{=}o$ and $\delta{=}o$}\\
       B_x        &\text{: $\alpha{=}o$ and $\delta{=}u$}\\
       0          &\text{: otherwise}\\
     \end{cases}
  \label{general-B-solution}
\end{align}
where $\nu{=}e$ if $x$ is an i-type boundary and $\nu{=}u$ if $x$ is
an n-type boundary.

Note in particular that any bulk field can couple to a $u$--field on
an n-type boundary, but for an i-type boundary an even bulk field can
only couple to an even boundary field and an odd bulk field to an odd
boundary field. This indicates that the $\Z_2$--symmetry of the
boundary structure constants for i-type boundaries carries
over to the bulk.

\subsection{Bulk structure constants}

The bulk structure constants can now be obtained from
\eqref{sewing:BBb}. Take $a{=}\o$, the sum on the l.h.s.\ then reduces to
the term where $\B{\o}{m_\rho}{k_\gamma}{\neq}0$, i.e.\ $k_\gamma{=}\o_\rho$. 
On the r.h.s.\ the sum over ${p_\nu}$ and ${q_\eps}$ reduces
to ${p_\nu}{=}\o_\alpha$, ${q_\eps}{=}\o_\beta$. The $r$--sum is also
reduced to one element. We are left with:
\begin{align}
  &\C{i_\alpha}{j_\beta}{m_\gamma} = 
     \exp\left(i\tfrac\pi{2}(h_{\o_\gamma}{+}h_{\o_\alpha}{-}h_{\o_\beta}{+}2(h_j{-}h_r)
        {+}h_m{-}\hb_m{-}h_i{+}\hb_i{-}h_j{+}\hb_j)\right)
     \notag\\
  &\cdot
     \frac{\B{\o}{i_\alpha}{\o_\alpha} 
        \B{\o}{j_\beta}{\o_\beta} }{ \B{\o}{m_\gamma}{\o_\gamma} }
     \cdot
     \F{\o_\alpha}{\mu_u}{\mu_u}{\o_\beta}{\mu_u}{\o_\gamma}
     \F{\o_\alpha}{\o_\gamma}{\jb}j{\o_\beta}r 
     \F i{\ib}rj{\o_\alpha}m 
     \F m{\ib}{\jb}{\o_\gamma}r{\bar m}
  \label{general-C-formula}
\end{align}
where $r{=}j$ if $\alpha{=}e$ and $r{=}j^*$ if $\alpha{=}o$. The first
F-matrix element in \eqref{general-C-formula} implements an even/odd
coupling rule for bulk fields, i.e.\ the only combinations for 
$\{\alpha,\beta,\gamma\}$ which can be nonzero are $\{e,e,e\}$ and
$\{e,o,o\}$.

This implies that apart from other symmetries the bulk structure
constants may have there is a manifest $\Z_2$--symmetry which sends
even bulk fields to themselves and odd one to minus themselves. Together
with section \ref{Collection-of-results} and
eqn.~\eqref{general-B-solution} we can now conclude that any
correlator with no or only i-type boundaries is manifestly
invariant under the $\Z_2$--symmetry $e\rightarrow e$ and
$o\rightarrow -o$ applied to bulk and boundary simultaneously.

It is useful to know the bulk two-point functions in the present
normalisation of fields. From the explicit expression in
\eqref{general-C-formula} we get:
\begin{align}
  &\C{i_\alpha}{i_\alpha}1 = \frac{\S 1i}{\S 11} &&
  \text{for diagonal fields $i_\alpha$ (i.e.\ $i{=}\ib$)} \notag\\
  &\C{i_o}{i_o}1 = A 
    \cdot \F{\ib}{\o_o}{\o_o}{\ib}{i}{1} 
    \cdot \frac{\S 1i}{\S 11} \cdot 
    (-1)^{s(i_o)}&&
  \text{for nondiagonal fields $i_o$} 
  \label{bulk-2pt-functions}
\end{align}
where $s(i_\alpha)=h_i{-}\hb_i\in\Z$ is the spin of the bulk field
$\phi_{i_\alpha}$.

In~\cite{PeZ95} it was shown that the minimal model bulk
structure constants in the  A-- and D--series
are related by rational numbers, called relative structure
constants. Taking the explicit expression~\eqref{general-C-formula}
together with the normalisation~\eqref{bulk-2pt-functions} we find
(numerical) agreement with these results. Also, for unitary models the
signs of the two-point structure constants
in~\eqref{bulk-2pt-functions} are the same as in \cite{PeZ95}, where
they were shown to lead to real bulk structure constants (in the
unitary case).

Note that if all bulk fields are even the solution
\eqref{general-C-formula} takes the same form as in the A-series:
\begin{align}
  \C{i_e}{j_e}{m_e} &= \left( \F jiijm1 \right)^{-1}
  \label{bulk-C-for-even-fields}
\end{align}
Together with equations \eqref{bsc-only-even-fields} for the boundary
structure  constants and
\eqref{general-B-solution} for the bulk-boundary couplings this illustrates
another interesting point. If one considers the even fields alone,
that is all bulk and boundary fields that are invariant under the
$\Z_2$--symmetry mentioned above, they form a subalgebra (as the
coupling $e,e\rightarrow o$ is not allowed) and the structure
constants are identical to the A-series in the following way:

Consider the A-series boundary theory associated to the pair of
diagrams $(A_{p-1},A_{q-1})$. In this theory consider only the
boundary conditions $(\alpha,\beta)$ with 
$\alpha<\frac p2$ and $\beta<\frac q2$ or
$\alpha>\frac p2$ and $\beta>\frac q2$ (compare to
\eqref{even-odd-weights}) and the bulk fields $i{\otimes}i$ where $i$
has Kac-labels $(r,s)$ with both $r$ and $s$ odd. Then this is a
closed subset of fields of the A-series theory and in the
normalisation chosen its
structure constants coincide with those of the invariant fields of the
D-series theory $(A_{p-1},D_{\frac q2 +1})$.
This correspondence seems natural form the point of view that the
invariant part of the D-diagram is an A-diagram.

\subsection{Real structure constants}
\label{Real-structure-constants}

In the normalisation chosen in this paper the F-matrices are all
real. This implies first of all that the boundary structure constants
as given in section~\ref{Collection-of-results} are all real. 

For the bulk structure constants we have seen in
\eqref{bulk-C-for-even-fields} that they are real as 
long as only even fields are involved. If odd fields are present
consider \eqref{general-C-formula} in the case
$\C{i_o}{j_o}{m_e}$. The phase factors cancel and we are left with a
real expression. All other cases are real as well, as can for example
be seen from the three-point function 
$\C{i_\alpha}{j_\beta}{k_\gamma}\C{k_\gamma}{k_\gamma}1{=}%
\C{j_\beta}{k_\gamma}{i_\alpha}\C{i_\alpha}{i_\alpha}1$ 
together with the fact that all two-point structure constants
\eqref{bulk-2pt-functions} are real. Thus in the present normalisation
all bulk structure constants are real, both for unitary and
non-unitary models.

For the bulk-boundary couplings in \eqref{general-B-solution} we can
apply the identity \eqref{F-identity-conjugate} to obtain the result:
\begin{align}
  \left( \B x{(i{\otimes}\ib)_\alpha}{\l_\delta} \right)^* &= 
  \B x{(\ib{\otimes}i)_\alpha}{\l_\delta}
  \label{B-conjugation}
\end{align}
This means that complex conjugation relates the bulk-boundary
coupling for a bulk field of spin $s$ to that of the field of spin
$-s$. In particular all bulk-boundary couplings for diagonal bulk
fields are real.

It is in general not possible to choose a basis of primary fields
s.t.\ all structure constants are real.
To see this we construct a gauge invariant expression from the sewing
constraints which cannot be fulfilled by real structure constants.
Consider for example two diagonal bulk fields $i_\alpha$ and $m_\rho$
and a field $j_\beta$ with spin one. Take \eqref{sewing:BBb} with $k{=}1$
and $a{=}\o$. This forces $p_\nu{=}q_\eps{=}\o_o$ and $r=\jb$. We get:
\begin{align}
   \sum_{\rho} \C{i_\alpha}{j_\beta}{m_\rho} \B{\o}{m_\rho}{1}
   &= e^{i\tfrac{\pi}2} \cdot \F ii{\jb}j{\o_o}m \cdot
   \B{\o}{i_\alpha}{\o_o}\B{\o}{j_\beta}{\o_o}\bc{\o}{\o}{\o}{\o_o}{\o_o}1 
\end{align}
The F-matrix entry is real and will in general be nonzero, provided
that all fusions are allowed. As discussed in
section~\ref{Bulk-field-content} the spin one field $j_\beta$ has to
couple to $\o_o$ (i.e.\ $\B{\o}{j_\beta}{\o_o}{\neq}0$) and we can find a
diagonal field $i_\alpha$ that couples to $\o_o$. The boundary
two-point function has 
to be nonzero by assumption b) in section~\ref{Boundary-structure-constants}.
Hence the r.h.s.\ is nonzero and thus 
some of the D-series structure constants will have nonzero imaginary
part if they solve the sewing constraints.

We can however consider the following non-primary basis of bulk fields to
obtain real structure constants: Diagonal (even or odd) bulk fields
stay as they are and for a pair $\phi_s$, $\phi_{-s}$ of nondiagonal
bulk fields with spins $\pm s$ we define a new set of fields as follows:
\begin{align}
  \phi^r = \tfrac 12 \left( \phi_s + \phi_{-s} \right)
  \qquad&\qquad
  \phi^i = \tfrac{1}{2i} \left( \phi_s - \phi_{-s} \right)
\end{align}
The new fields $\phi^r$ and $\phi^i$ are no longer primary, but one
finds that the
coefficients describing their behaviour under arbitrary conformal
mappings are all real. Since the OPE is determined from the
transformation behaviour of the fields one expects the coefficients
appearing in the OPE to be real as well. 

One can see explicitly from \eqref{B-conjugation} that all
bulk-boundary couplings $\B x{\phi^r}{\psi}$ and  $\B x{\phi^i}{\psi}$
are real and one can verify numerically that all bulk structure
constants are real in the new basis as well. This holds for both
unitary and non-unitary models.

\subsection{Discrete symmetries of the structure constants}
\label{Discrete-symmetries-of-the-structure-constants}

Denote the set of bulk and boundary structure constants given in 
\eqref{collection-first-bsc-formula}--\eqref{collection-last-bsc-formula}
and
\eqref{bulk-field-normalisation}--\eqref{general-C-formula}
with~$A$.
We see that $A$ depends explicitly on the
assignment of $e/o$--labels in section~\ref{The-D-series} and the
ordering of the boundary conditions in
section~\ref{Ordering-of-boundary-conditions}.

Denote the set of structure constants obtained by the same procedure,
but with a different initial assignment of $e/o$--labels in
section~\ref{The-D-series} as $B$.
It was shown in this paper
that any solution to the sewing constraints with the D-series field
content can be brought to the form $A$.
In particular there is a change of basis in the set of primary fields
that maps $B$ to $A$.

For the primary fields in $A$ we considered a $\Z_2$--action that
takes even fields to themselves and odd fields to minus
themselves. This action leaves the structure 
constants involving no or only i-type boundaries invariant.

However a different initial
$e/o$--labelling leads to a different action of $\Z_2$ on the
boundary fields. The $\Z_2$--action of $B$ does via the change of
basis induce a $\Z_2$--action on $A$ which is different from the
original one. It follows that each distinct
$e/o$--assignment results in a distinct $\Z_2$--action on $A$.

Going back to section~\ref{The-D-series} we see
that for each i-type boundary $x{\neq}\o$ we are free to choose which
of the two fields living on $\bfb{\o}{}x$ we call $x_e$ and which
$x_o$. The specific assignment made in section~\ref{The-D-series} was
motivated on physical grounds.

There are altogether $N{=}2^{(p-1)(q-2)/4-1}$
distinct assignments. Thus among other symmetries the
structure constants may have, there are $N$ $\Z_2$'s acting
on the set of primary fields that leave the subset of the structure
constants in $A$ that involve no or only i-type boundary conditions
invariant. One can convince oneself that the $\Z_2$'s
all act in the same way on the bulk fields but can be distinguished
from their action on the boundary fields. The collection of $\Z_2$'s
just presented does not necessarily exhaust all discrete symmetries as
can be seen for example in the three-states Potts model which has an
additional $\Z_3$--symmetry of whose $\Z_2$ subgroups only one is
contained in the above list.

\subsection{Disc partition function and Pasquier Algebra}
\label{Disc-partition-function-and-Pasquier-Algebra}

Up two now all constraints on the structure constants were derived in
such a way that the vacuum expectation values $\bnpt 1x{UHP}$ cancelled. We will
next derive some constraints for the vacuum expectation values of the unit
disc $\bnpt 1x{D}$. Consider the two-point function 
$\npt{\bp xy{i_\alpha}\bp yx{i_\alpha}}$ on the unit disc. Taking
different limits leads to the constraint:
\begin{align}
  \bc xyx{i_\alpha}{i_\alpha}1 \bnpt 1x{D} &=
  \bc yxy{i_\alpha}{i_\alpha}1 \bnpt 1y{D}
  \label{different-bnd-0pt-fn-relation}
\end{align}
The vacuum expectation values $\bnpt 1x{D}$, $\bnpt 1y{D}$
can be interpreted as the partition
function of the unit disc with different boundary conditions. 
This expression thus relates the partition functions of different
boundary conditions in the same geometry. The relation
\eqref{different-bnd-0pt-fn-relation} remains true for
different (genus zero) geometries, but the actual values of the
partition functions do depend on the shape of the system
(see e.g.~\cite{CaP88}). For example the partition function for a disc of
radius~$r$ is given by $\bnpt 1x{D} \cdot r^{c/6}$, where $c$ is
the central charge \cite{CaP88}.

Substituting the explicit boundary two-point functions into
\eqref{different-bnd-0pt-fn-relation} 
we arrive at the following expressions (let $x,y$ be i-type boundaries
and $a,b$ be n-type boundaries):
\begin{align}
  \frac{\bnpt 1x{D}}{\S 1{x_e}} = \frac{\bnpt 1y{D}}{\S 1{y_e}} \qquad\qquad
  \frac{\bnpt 1x{D}}{\S 1{x_e}} = 2\cdot\frac{\bnpt 1a{D}}{\S 1{a_u}} \qquad\qquad
  \frac{\bnpt 1a{D}}{\S 1{a_u}} = \frac{\bnpt 1b{D}}{\S 1{b_u}}
  \label{0pt-constraints}
\end{align}

In \cite{BPZ98} the matrices $n_i$ that descibe the boundary field
content are simultaneously diagonalised in the following way:
\begin{align}
  \n ixy &= \sum_{j_\beta \text{ diag.}} \frac{\S ij}{\S 1j} 
  \cdot \psi_{x,j_\beta} (\psi_{y,j_\beta})^*
  \label{BPPZ-n-relation}
\end{align}
where the sum runs over all (even or odd) diagonal bulk fields and the
vectors $\psi_x$ form a complete orthonormal basis.

Replicating the line of argument in section~8 of \cite{Run99} in the
present notation we get the following expression for the matrices
$n_i$:
\begin{align}
  \n ixy &= \sum_{j_\beta \text{ diag.}} \frac{\S ij}{\S 1j} 
  \cdot \frac{\S 1j}{\C{j_\beta}{j_\beta}1 \npt{0|0}} 
  \B{x}{j_\beta}1 \bnpt 1x{D}
  \B{y}{j_\beta}1 \bnpt 1y{D}
\end{align}
Using \eqref{bulk-2pt-functions} and normalising the bulk vacuum: 
\begin{align}
  \npt{0|0} &= \S 11
\end{align}
we find the following relation between the bulk-boundary couplings and
the vectors $\psi_x$:
\begin{align}
  \psi_{x,j_\beta} &= \B{x}{j_\beta}1 \bnpt 1x{D}
  \label{first-psi-B-relation}
\end{align}
By evaluating
\eqref{first-psi-B-relation} for $j{=}1$ and recalling that 
$\B{x}11{=}1$ we get:
\begin{align}
  \bnpt 1x{D} = \psi_{x,1}
  \label{psi-to-0pt-relation}
\end{align}
Furthermore we see that the the bulk-boundary couplings to the
identity on the boundary are given by a ratio of $\psi$'s \cite{BPZ98}:
\begin{align}
  \B{x}{j_\beta}1  = \frac{\psi_{x,j_\beta}}{\psi_{x,1}}
  \label{B-is-ratio-of-psis}
\end{align}
In the present normalisation the $B$'s form one dimensional
representations of an algebra, a fact that was used to classify
possible boundary conditions in
\cite{CaL91,PSS95,FuS97,BPZ98}. The algebra was 
identified as the Pasquier algebra \cite{Pas87} in \cite{BPZ98}:
\begin{align}
  \B{x}{i_\alpha}1 \B{x}{j_\beta}1 &=
  \sum_{k_\gamma} {M_{i_\alpha j_\beta}}^{k_\gamma} \cdot \B{x}{k_\gamma}1 
\end{align}
where
${M_{i_\alpha j_\beta}}^{k_\gamma}=\sum_x 
\psi_{x,i_\alpha}\psi_{x,j_\beta}(\psi_{x,k_\gamma})^*/\psi_{x,1}$.

From the relations \eqref{psi-to-0pt-relation}, \eqref{B-is-ratio-of-psis} and
\eqref{0pt-constraints} together with the constraint that the vectors
$\psi_x$ form an orthonormal basis we get the following expression:
\begin{align}
  \psi_{x,i_\alpha}&=\B{x}{i_\alpha}1 \psi_{x,1} &
  \psi_{x,1} = \begin{cases}
    \sqrt{2} \cdot \S 1{x_e} &\text{if $x$ is of i-type} \\
    \frac{1}{\sqrt{2}} \cdot \S 1{x_u} &\text{if $x$ is of n-type}
  \end{cases}
\end{align}
Since the bulk-boundary couplings are known from
\eqref{general-B-solution} one can now check that the $\psi$'s so
defined are all real and 
verify numerically that they fulfill \eqref{BPPZ-n-relation}, with 
$\n ixy$ on the l.h.s.\ taken from section \ref{The-D-series}.

\section{Conclusion}

The aim of this paper was to find all structure constants of the D-series
minimal model boundary conformal field theory and to point out some of
their properties. 

The structure constants can be calculated in several steps.
The basic ingredient are the fusion matrices, as all structure
constants are expressed in 
terms of these. A procedure to obtain the fusion matrices recursively
is given in \cite{Run99}, together with references to the original
literature. The matrices can also be extracted by
rescaling explicit expressions found e.g.\ in \cite{FusXX} to match
the normalisation fixed in appendix~A. As a
first step the D-series field content of the boundary theory is computed, as
described in section \ref{The-D-series}. Next all boundary
structure constants are obtained from the results collected in section
\ref{Collection-of-results}. The boundary theory is now fully
determined. The extension from the boundary to the bulk recovers the
familiar D-series bulk field content as the maximal set of bulk fields
consistent with the boundary theory (see section \ref{Bulk-field-content}).
Last, the remaining two sets of structure constants, the bulk-boundary
couplings and the bulk structure constants are given in
\eqref{bulk-field-normalisation}, \eqref{general-B-solution} and
\eqref{general-C-formula}.  

The boundary structure constants
$\bc xyz{r_\alpha}{s_\beta}{t_\gamma}$ and the bulk structure
constants $\C{i_\alpha}{j_\beta}{k_\gamma}$ obtained by this procedure
are all real, both for unitary and non-unitary minimal models.
However some of the bulk-boundary couplings
$\B x{i_\alpha}{\l_\delta}$ are imaginary. There is in general no
basis of primary fields such that all structure constants real. In
section~\ref{Real-structure-constants} a non-primary basis of fields
was presented in which all three sets of structure constants are real.

The solution described above was obtained by deriving a necessary form
of the structure constants. It was shown that any solution to the
sewing constraints \eqref{sewing:bbbb}--\eqref{sewing:BBBB} for the
D-series field content can 
be brought to this form. So just as for the A-series it follows
that the structure constants are unique up to redefinition of the
fields. It was however not analytically shown that the given structure
constants do indeed solve the full set of genus zero sewing
constraints, since they are overdetermined and only a subset was used
for the actual solution. But numerical tests carried out for several
minimal models where each of the equations
\eqref{sewing:bbbb}--\eqref{sewing:BBBB} was checked repeatedly with a
different random selection of fields show no contradiction.

It was also found that a subset of the D-series structure constants
have a manifest $\Z_2$--symmetry: If no or only i-type boundaries
are present in a correlator it is invariant under the change
`even'$\rightarrow$`even' and `odd'$\rightarrow -$`odd'.
This symmetry can be associated to the $\Z_2$--symmetry of the
D-diagram in the sense that the boundary conditions fall in two
classes: the i-type boundaries are associated to nodes left
invariant by that symmetry, and the n-type boundaries to nodes which
are not. We have also
seen that the even fields alone form a closed subalgebra with structure
constants equal to the corresponding ones of an A-type theory.

In deriving the structure constants we have used only the Virasoro
algebra, and not considered the fact that it could be part of a
larger symmetry algebra, as is the case for
$D_{\text{even}}$~models. From the present point of view, curious
properties of the solution such as the boundary field content, the
even/odd coupling rule or numerical coincidences in the structure
constants, are obtained as results of the calculation, but lack an 
explanation in terms of the additional structure present in the
theory. One can use the maximally extended chiral algebra as a
starting point for the analysis of a boundary conformal field theory
(see e.g.~\cite{SymXX} and references therein) and it would be
interesting to interpret the explicit results presented here in this
more general framework.

In section~\ref{Discrete-symmetries-of-the-structure-constants} it was
pointed out that an
initial freedom in the labelling of the boundary fields actually
implies that the just mentioned $\Z_2$ is part of a collection
of $2^{(p-1)(q-2)/4-1}$ $\Z_2$'s, each acting in a distinct way on the
set of primary boundary fields. This action leaves
structure constants with no or only i-type boundaries invariant. 

Finally the vacuum expectation values $\bnpt 1x{D}$ of the unit disc with
different boundary conditions were computed and it was pointed out
that in the present normalisation the bulk-boundary couplings 
$\B x{i_\alpha}1$ form one-dimensional representations of the Pasquier
algebra.

To make contact with different conventions in the literature the bulk
and boundary fields may have to be redefined. The normalisations
chosen in this paper can be obtained from \eqref{bulk-2pt-functions}
in terms of the bulk two-point functions, or from
\eqref{bulk-field-normalisation} in terms of the bulk-boundary couplings to
the identity field on the boundary.

\vspace{10pt}
\noindent{\bf Acknowledgements} --- This work was supported by the
EPSRC, the DAAD and King's College London. I am especially grateful to
my supervisor G.M.T.~Watts for suggesting the project, encouragement,
criticism and helpful conversations. Furthermore I wish to thank
V.B.~Petkova, P.~Ruelle and J.-B.~Zuber for useful discussions.

\section*{Appendix A: Derivation of the five-point constraint}

This appendix was included to illustrate how the notation and
techniques for calculating with conformal blocks described in
\cite{MSb90} were used to rederive the sewing constraints in
\cite{Lew92}.

The sewing constraints arise from taking two different limits in a given  
correlator where the fields approach each other or the boundary and
demanding that the two sets of structure constants and
conformal blocks give rise to the same function.

Conformal blocks associated to different asymptotic regimes can be
transformed into each other using the fusion-- or F--matrices. The
F--matrices are defined as follows:
\begin{align}
     \bL{\cbB ijk{\l}p{(z)}{(w)}} &= \sum_q \F ijk{\l}pq
     \bL{\cbC ijk{\l}q{(w)}{(z-w)}}
     \label{A:F-matrix-definition}
\end{align}
The conformal blocks appearing in \eqref{A:F-matrix-definition} solve
the linear differential equations associated with the (chiral)
correlator $\os i \ph j(z) \ph k(w)\is l$.

The normalisation convention for conformal blocks chosen in this paper
is (the dots stand for a regular expression in $w$):
\begin{align}
     &\bL{\cbB ijk{\l}p{(1)}{(w)}} = w^{h_p-h_k-h_\l}(1+\cdots) \notag\\
     &\bL{\cbC ijk{\l}q{(w)}{(1-w)}} = (1-w)^{h_q-h_j-h_k}(1+\cdots)
\end{align}

The example we consider is the derivation of the constraint resulting
from taking different limits in the correlator involving two bulk
fields $\ph{i_\alpha}$, $\ph{j_\beta}$ and one boundary field 
$\bp aa{k_\gamma}$ on the upper half plane with the boundary condition
labelled $a$. 

In the first limit we take the two bulk fields to the boundary
and are left with a three-point function on the boundary. In the
second limit we start by taking the OPE of the two bulk fields and
then take the remaining bulk field to the boundary. Let $z{=}x_z{+}iy_z$
and $w{=}x_w{+}iy_w$. The asymptotic
behaviour in the two limiting cases is then given by:
\begin{align}
     &\npt{\ph {i_\alpha}(z,\zb) \ph {j_\beta}(w,\wb) \bp aa{k_\gamma}(x)}
     \notag\\
     &\quad\simuu {y_z}0{y_w}0 \sum_{p,q} 
          \left(\sum_{\nu,\eps} \B a{i_\alpha}{p_\nu} \B a{j_\beta}{q_\eps} 
          \bc aaa{p_\nu}{q_\eps}{k_\gamma} \bc aaa{k_\gamma}{k_\gamma}1
          \bnpt 1a{UHP} \right)
          \notag\\
          &\qquad \cdot (2y_z)^{h_p-h_i-\hb_i}
          (2y_w)^{h_q-h_j-\hb_j}(x_z-x_w)^{h_k-h_p-h_q} \notag \\
          &\qquad\qquad \cdot (x_z-x)^{h_q-h_p-h_k} (x_w-x)^{h_p-h_k-h_q}
          \notag\\
     &\quad\simuu zw\zb\wb
     \sum_{m,\bar m} 
          \left(\sum_{\rho} \C {i_\alpha}{j_\beta}{m_\rho} \B a{m_\rho}{k_\gamma}
          \bc aaa{k_\gamma}{k_\gamma}1 \bnpt 1a{UHP} \right)
          \notag\\
          &\qquad \cdot (z-w)^{h_m-h_i-h_j} (\zb-\wb)^{\hb_m-\hb_i-\hb_j}
          (2y_w)^{h_k-h_m-\hb_m} (x_w-x)^{-2h_k}
     \label{app:limits-in-corr}
\end{align}
On the other hand the correlator can be expressed as a linear
combination of conformal blocks. We use two sets of conformal blocks
to express the correlator in two different ways, one associated to
each asymptotic behaviour in \eqref{app:limits-in-corr}. We obtain the
following two linear combinations: 
{\allowdisplaybreaks
\begin{align}
     &\npt{\ph{i_\alpha}(z,\zb) \ph {j_\beta}(w,\wb) \bp aa{k_\gamma}(x)} \notag\\
     &\qquad =  \sum_{p,q} c_{p,q} \cdot
\bL{\blockA 0{}{\zb}\blockUp{\blockD pi{z-\zb}\blockB{\ib}}%
\blockE\blockB p\cbC{}{j}{\jb}{}q{\wb}{w-\wb}\cbA{}k0x}\notag\\
     &\qquad = \sum_{m,\bar m} d_{m,\bar m} \cdot 
\bL{\blockA 0{}{w}\blockUp{\blockD k{}{w-\wb}\blockUp{\cbA mij{z-w}}%
\blockE\blockB{\bar m}\blockA{}{\ib}{\zb-\wb}\blockB{\jb}}%
\blockE\blockE\blockE\blockE\blockE\blockB k\blockE\blockE\blockE%
\blockE\cbA{}k0x}
     \label{app:conf-block-sum}
\end{align}}
Taking the limits calculated in \eqref{app:limits-in-corr} for the
exact expressions in terms of conformal blocks
\eqref{app:conf-block-sum} relates the coefficients $c_{p,q}$ and
$d_{m,\bar m}$ to the products of structure constants obtained by
applying the OPE. The precise relation is:
\begin{align}
     c_{p,q} \cdot e^{i\frac{\pi}{2}(h_p+h_q-h_i-\hb_i-h_j-\hb_j)} 
     &= \sum_{\nu,\eps} \B a{i_\alpha}{p_\nu} \B a{j_\beta}{q_\eps} 
     \bc aaa{p_\nu}{q_\eps}{k_\gamma} \bc aaa{k_\gamma}{k_\gamma}1
     \bnpt 1a{UHP}\notag\\
     d_{m,\bar m} \cdot e^{i\frac{\pi}{2}(h_k-h_m-\hb_m)}
     &= \sum_{\rho} \C {i_\alpha}{j_\beta}{m_\rho} \B a{m_\rho}{k_\gamma}
     \bc aaa{k_\gamma}{k_\gamma}1 \bnpt 1a{UHP}
     \label{app:conf-block-coeff}
\end{align}
The phase factors originate from relating $z-\zb = 2i y_z$ to $2y_z$ etc.
The sums over $\nu,\eps$ and $\rho$ in \eqref{app:conf-block-coeff}
take care of fields with 
multiplicities. Fields that transform in the same representation of the
Virasoro algebra show the same asymptotic behaviour and cannot be
discriminated by conformal blocks. Hence their structure constants
occur as a sum in front of the according conformal block.

The two sets of conformal blocks in \eqref{app:conf-block-sum} are
related by a basis transformation. This transformation can be carried
out in several steps making use of two basic moves of braiding and
fusion implemented by the B-- and F-- matrix (see \cite{MSb90} for
details). One possible way to perform the basis transformation is as
follows: 
{\allowdisplaybreaks
\begin{align}
&\bL{\blockA 0{}{\zb}\blockUp{\blockD pi{z-\zb}\blockB{\ib}}%
\blockE\blockB p\cbC{}{j}{\jb}{}q{\wb}{w-\wb}\cbA{k}k0x}
\notag\\
& = \sum_r \left(\F pj{\jb}k{}{} \right)^{-1}_{qr}\cdot
\bL{\blockA 0iz\blockC i{\ib}\zb\blockC pjw%
\blockC r{\jb}\wb\blockC kkx\blockB 0}
\notag\\
& = \sum_{r,m} \F pk{\jb}jqr \Bmat-i{\ib}jrpm\cdot
\bL{\blockA 0iz\blockC ijw\blockC m{\ib}\zb%
\blockC r{\jb}\wb\blockC kkx\blockB 0}
\notag\\
& = \sum_{r,m,\bar m} 
\F pk{\jb}jqr \Bmat-i{\ib}jrpm \F m{\ib}{\jb}kr{\bar m}
\cdot\bL{\blockA 0{}w\blockUp{\blockD mi{z-w}\blockB j}%
\blockE\blockB m\cbC{}{\ib}{\jb}{}{\bar
m}{\wb}{\zb-\wb}\cbA{k}k0x}
\notag\\
& = \sum_{r,m,\bar m} e^{-i\pi(h_i+h_r-h_p-h_m)}
\F pk{\jb}jqr \F i{\ib}rjpm \F m{\ib}{\jb}kr{\bar m}\notag\\
&\qquad\cdot
\bL{\blockA 0{}{w}\blockUp{\blockD k{}{w-\wb}\blockUp{\cbA mij{z-w}}%
\blockE\blockB{\bar m}\blockA{}{\ib}{\zb-\wb}\blockB{\jb}}%
\blockE\blockE\blockE\blockE\blockE\blockB k\blockE\blockE\blockE%
\blockE\cbA{}k0x}
\label{app:5pt-block-xfer}
\end{align}
}
Putting \eqref{app:conf-block-sum}, \eqref{app:conf-block-coeff}
and \eqref{app:5pt-block-xfer} together we recover the sewing
constraint \eqref{sewing:BBb}:
\begin{align}
     &\sum_{\rho} \C {i_\alpha}{j_\beta}{m_\rho} \B a{m_\rho}{k_\gamma}
     \bc aaa{k_\gamma}{k_\gamma}1 \bnpt 1a{UHP}\notag\\
     &\qquad = \sum_{p,q}
     \left(\sum_{\nu,\eps} \B a{i_\alpha}{p_\nu} \B a{j_\beta}{q_\eps} 
     \bc aaa{p_\nu}{q_\eps}{k_\gamma} \bc aaa{k_\gamma}{k_\gamma}1
     \bnpt 1a{UHP}\right)\notag\\
     &\qquad\qquad \cdot \sum_{r}
     e^{i\frac\pi{2}(h_k+h_p-h_q-2h_r+h_m-\hb_m-h_i+
     \hb_i+h_j+\hb_j)}\notag\\
     &\qquad\qquad\qquad \cdot
     \F pk{\jb}jqr \F i{\ib}rjpm \F m{\ib}{\jb}kr{\bar m}
\end{align}
The summation range for $r$ is principally over all entries in the
Kac-table (quotiened by $\Z_2$) but the F--matrix entries will only be
nonzero if the following three fusions are allowed:
\begin{align}
   \bL{\cbA rjp{}}        \qquad 
   \bL{\cbA r{\jb}k{}} \qquad 
   \bL{\cbA r{\ib}m{}}
\end{align}

\section*{Appendix B: Some F-matrix identities}

The following constitutes a collection of F-matrix identities for
Virasoro minimal models used in the paper. They are either directly taken from
\cite{MSb90} or special cases thereof. Note that for the S--matrix we
have $\S 1i{=}\S 1{i^*}$ where the $*$-operation was defined
as $i^*=(r,q-s)$ if $i$ has Kac-labels $(r,s)$. 

{\allowdisplaybreaks
\begin{alignat}{2}
  & \Bmat \epsilon ijk\l pq = e^{i\pi\epsilon (h_i+h_\l-h_p-h_q)}
      \F ijlkpq\\
  & \F ijk\l pq = \F ji\l kpq = \F k\l ijpq \label{F-identity-first}\\
  & \sum_r \F abcdpr \F adcbrq = \delta_{p,q}
    && \F a{x_i}{\o_o}b{b^*}{{x_i}^*} \F ab{\o_o}{x_i}{{x_i}^*}{b^*} = 1
    \label{F-identity-b} \\
  & \F iijj1k \F jiijk1 = \frac{\S 11\cdot \S 1k}{\S 1i\cdot\S 1j}
    && \F{x_e}{x_e}{\o_o}{\o_o}{1}{x_o} \cdot \F{x_e}{\o_o}{\o_o}{x_e}{x_o}{1} = 1
    \label{F-identity-a}  \\
  & \F jiijk1 = \frac{\S 1k}{\S 1j} \cdot \F ikkij1
    && \F{x_e}{\o_o}{\o_o}{x_e}{x_o}{1} = \F{x_o}{\o_o}{\o_o}{x_o}{x_e}{1} \\
  & \F iiii11 = \frac{\S 11}{\S 1i}
    && \F{\o_o}{\o_o}{\o_o}{\o_o}{1}{1} = 1 \\
  & \F njk\l pi \F \l ii\l n1 =\F \l ijpnk \F \l kk\l p1 
  \label{F-identity-last}
\end{alignat}
\begin{alignat}{1}
  & \sum_{s} e^{i\pi(h_p+h_q+2h_i-2h_s-\frac 12 (h_k+h_\l))}
      \F i{\bar i}pqks \F iqp{\bar i}s{\l} \notag\\
  & \hspace{2cm}  = \sum_m e^{i\pi(-h_p-h_q-2\hb_i+2h_m+\frac 12 (h_k+h_\l))}
  \F qpi{\bar i}km \F ipq{\bar i}m{\l} 
  \label{F-identity-conjugate}
\end{alignat}
}



\begin{thebibliography}{99}
\small
\raggedright
\bibitem{BPZ84}
    A.A.~Belavin, A.M.~Polyakov, A.B.~Zamolodchikov,
       {\em Infinite conformal symmetry in two-dimensional quantum
            field theory},
       Nucl.~Phys.~B241 (1984) 333--380.
\\[-20pt]
\bibitem{YBk}
    Ph.~Di Francesco, P.~Mathieu, D.~S\'en\'echal,
       {\em Conformal Field Theory},
       Springer 1998.
\\[-20pt]
\bibitem{DF84}
    Vl.S.~Dotsenko, V.A.~Fateev,
       {\em Conformal algebra and multipoint correlation functions in
            2d statistical models},
       Nucl.~Phys.~B240 (1984) 312--348.
       {\em Four-point correlation functions and the operator agebra
            in 2d conformal invariant theories with central charge $c{\le}1$},
       Nucl.~Phys.~B251 (1985) 691--734.
\\[-20pt]
\bibitem{CIZ87}
    A.~Capelli, C.~Itzykson, J.-B.~Zuber,
       {\em Modular invariant partition functions in two dimensions},
       Nucl.~Phys.~B280 (1987) 445--465.
\\[-20pt]
\bibitem{Car84}
    J.L.~Cardy,
       {\em Conformal invariance and surface critical behavior},
       Nucl.~Phys.~B240 (1984) 514--532.
\\[-20pt]
\bibitem{CaL91}
    J.L.~Cardy, D.C.~Lewellen,
       {\em Bulk an boundary operators in conformal field theory},
       Phys.~Lett.~B259 (1991) 274--278.
\\[-20pt]
\bibitem{PSS95}
    G.~Pradisi, A.~Sagnotti, Ya.~S.~Stanev, 
       {\em The open descendants of nondiagonal SU(2) WZW models} 
       Phys.~Lett.~B356 (1995) 230--238, {\tt hep-th/9506014}.
       {\em Completeness conditions for boundary operators in 2-D
            conformal field theory},
       Phys.~Lett.~B381 (1996) 97--104, {\tt hep-th/9603097}.
\\[-20pt]
\bibitem{FuS97}
    J.~Fuchs, C.~Schweigert,
       {\em A Classifying algebra for boundary conditions},
       Phys.~Lett.~B414 (1997) 251--259, {\tt hep-th/9708141}.
\\[-20pt]
\bibitem{BPZ98}
    R.E.~Behrend, P.A.~Pearce, J.-B.~Zuber,
       {\em Integrable boundaries, conformal boundary conditions and
            A-D-E fusion rules},
       J.~Phys.~A31 (1998) L763-L770, {\tt hep-th/9807142}.
    R.E.~Behrend, P.A.~Pearce, V.B.~Petkova, J.-B.~Zuber,
       {\em On the Classification of Bulk and Boundary Conformal Field
            Theories},
       Phys.~Lett.~B444 (1998) 163--166, {\tt hep-th/9809097}.
\\[-20pt]
\bibitem{Lew92}
    D.C.~Lewellen, 
       {\em Sewing constraints for conformal field theories on
            surfaces with boundaries},
       Nucl.~Phys.~B372 (1992) 654--682.
\\[-20pt]
\bibitem{MSb90}
    G.~Moore and N.~Seiberg,
       {\em Lectures on RCFT},
       Physics, Geometry, and Topology, Plenum Press, New York, 1990.
\\[-20pt]
\bibitem{BPZ99}
    R.E.~Behrend, P.A.~Pearce, V.B.~Petkova, J.-B.~Zuber,
       {\em Boundary Conditions in Rational Conformal Field Theories},
       Nucl.~Phys.~B570 (2000) 525--589, {\tt hep-th/9908036}.
\\[-20pt]
\bibitem{StrXX}
    J.~Fuchs, A.~Klemm,
       {\em The computation of the operator algebra in nondiagonal
            conformal field theories}, 
       Annals~Phys.~194 (1989) 303.
    V.B.~Petkova,
       {\em Structure constants of the (A,D) minimal $c{<}1$ conformal
            models},
       Phys.~Lett.~B225 (1989) 357--362. 
\\[-20pt]
\bibitem{PeZ95}
    V.B.~Petkova, J.-B.~Zuber,
       {\em On structure constants of sl(2) theories},
       Nucl.~Phys.~B438 (1995) 347--372, {\tt hep-th/9410209}.
\\[-20pt]
\bibitem{Run99}
    I.~Runkel,
       {\em Boundary structure constants for the A-series Virasoro
            minimal models},
       Nucl.~Phys.~B549 (1999) 563--578, {\tt hep-th/9811178}.
\\[-20pt]
\bibitem{WalXX}
    W.~Eholzer, M.~Flohr, A.~Honecker, R.~Hubel, W.~Nahm, R.~Varnhagen,
       {\em Representations of W-algebras with two generators and new
            rational models},
       Nucl.~Phys.~B383 (1992) 249--290. 
    P.~Bouwknegt, K.~Schoutens,
       {\em W-symmetry in conformal field theory},
       Phys.~Rept.~223 (1993) 183--276, {\tt hep-th/9210010}.
\\[-20pt]
\bibitem{FusXX}
    L.~Alvarez-Gaum\'e, C.~Gomez, G.~Sierra,
       {\em Quantum group interpretation of some conformal field theories},
       Phys.~Lett.~B220 (1989) 142--152.
    G.~Felder, J.~Fr\"ohlich, G.~Keller,
       {\em Braid matrices and structure constants for minimal
            conformal models},
       Comm.~Math.~Phys.~124 (1989) 647--664.
    P.~Furlan, A.Ch.~Ganchev, V.B.~Petkova,
       {\em Fusion matrices and $c{<}1$ (quasi) local conformal
            field theories},
       Int.~J.~Mod.~Phys.~A5 (1990) 2721--2735.
\\[-20pt]
\bibitem{Car89}
    J.L.~Cardy,
       {\em Boundary conditions, fusion rules and the Verlinde formula},
       Nucl.~Phys.~B324 (1989) 581--596.
\\[-20pt]
\bibitem{Rue98}
    P.~Ruelle,
       {\em Symmetric boundary conditions in boundary critical phenomena},
       J.~Phys.~A32 (1999) 8831--8850, {\tt hep-th/9904100}.
\\[-20pt]
\bibitem{CaP88}
    J.L.~Cardy, I.~Peschel,
       {\em Finite size dependence of the free enery in
            two-dimensional critical systems},
       Nucl.~Phys.~B300 (1988) 377--399.
\\[-20pt]
\bibitem{Pas87}
    V.~Pasquier,
       {\em Operator content of the ADE lattice models},
       J.~Phys.~A20 (1987) 5707--5733.
\\[-20pt]
\bibitem{SymXX}
    J.~Fuchs, C.~Schweigert,
       {\em Symmetry breaking boundaries. 1.~General theory},
       Nucl.~Phys.~B558 (1999) 419--483, {\tt hep-th/9902132}.
       {\em Symmetry breaking boundaries. 2.~More structures: Examples},
       Nucl.~Phys.~B568 (2000) 543--593, {\tt hep-th/9908025}.
\end{thebibliography}
\end{document}